\documentclass[aps,prb,superscriptaddress,twocolumn,showpacs,preprintnumbers,amsmath,amssymb]{revtex4}
\usepackage{color}
\usepackage{array}
\usepackage{amsmath}
\usepackage{amssymb}
\usepackage{epsfig}
\usepackage{graphicx}
\usepackage{bbm}

\newcommand{\Ref}[1]{Ref.~\onlinecite{#1}}

\newcommand{\bss}{{\boldsymbol{\sigma}}}
\newcommand{\bst}{{\boldsymbol{T}}}
\newcommand{\bse}{{\boldsymbol{e}}}
\newcommand{\ie}{{\emph{i.e.~}}}
\makeatletter

\newcommand{\Rmnum}[1]{\expandafter\@slowromancap\romannumeral #1@}
\makeatother
\newcommand{\imth}{\hspace{1pt}\mathrm{i}\hspace{1pt}}

\newcommand{\eg}{{\emph{e.g.~}}}
\newcommand{\etc}{{\emph{etc.~}}}

\newcommand{\cs}{{C_6}}
\newcommand{\pxy}{{P_{xy}}}
\newcommand{\tk}{\textbf{k}}
\newcommand{\tq}{\textbf{q}}

\begin{document}
\title{Symmetry protected fractional Chern insulators and fractional topological insulators}

\author{Yuan-Ming Lu}\author{Ying Ran}
\affiliation{Department of Physics, Boston College, Chestnut Hill,
Massachusetts 02467, USA}
\date{\today}

\begin{abstract}
In this paper we construct fully symmetric wavefunctions for the spin-polarized fractional Chern insulators (FCI) and time-reversal-invariant fractional topological insulators (FTI) in two dimensions using the parton approach. We show that the lattice symmetry gives rise to many different FCI and FTI phases even with the same filling fraction $\nu$ (and the same quantized Hall conductance $\sigma_{xy}$ in FCI case). They have different symmetry-protected topological orders, which are characterized by different projective symmetry groups. We mainly focus on FCI phases which are realized in a partially filled band with Chern number one. The low-energy gauge groups of a generic $\sigma_{xy}=1/m\cdot e^2/h$ FCI wavefunctions can be either $SU(m)$ or the discrete group $Z_m$, and in the latter case the associated low-energy physics are described by Chern-Simons-Higgs theories. We use our construction to compute the ground state degeneracy. Examples of FCI/FTI wavefunctions on honeycomb lattice and checkerboard lattice are explicitly given. Possible non-Abelian FCI phases which may be realized in a partially filled band with Chern number two are discussed. Generic FTI wavefunctions in the absence of spin conservation are also presented whose low-energy gauge groups can be either $SU(m)\times SU(m)$ or $Z_m\times Z_m$. The constructed wavefunctions also set up the framework for future variational Monte Carlo simulations.
\end{abstract}


\maketitle


\section{Introduction}\label{INTRO}
Phases of matters in condensed matter systems can almost always be characterized by the Landau-Ginzburg symmetry breaking theory\cite{Landau1937,Landau1937a}. Experimental discovery of integer and fractional quantum Hall states in 2-D electron gas under a strong external magnetic field\cite{Klitzing1980,Tsui1982} has provided striking counter examples of this paradigm. The fractional quantum Hall liquids are particularly fascinating in the sense that their low energy excitations are quasi-particles carrying fractional electric charge\cite{Laughlin1983} and obeying anyonic statistics\cite{Arovas1984}. Although these liquid phases do not break physical symmetries, they are still different quantum phases. One measurable difference is their edge states: despite the fact that these liquids are all insulators in the bulk, they all possess certain edge metallic modes\cite{Wen1995}. In general different bulk phases host different edge states which can be detected by various experimental probes such as electric transport\cite{Wen1991a}.

A few years after the experimental discovery of integer quantum Hall effect(IQHE), Haldane showed that the essence of it is \emph{not} the external magnetic field\cite{Haldane1988}, by explicitly writing down a lattice model Hamiltonian of IQHE with zero net magnetic field. However, it takes more than two decades for people to show that similar statement is true even for FQHE. Recently results from a series of model studies\cite{Tang2011,Sun2011,Neupert2011,Sheng2011,Wang2011,Regnault2011,Neupert2011a,Xiao2011,Hu2011}, including convincing evidences from exact diagonalizations\cite{Neupert2011,Sheng2011,Wang2011,Regnault2011,Neupert2011a,Xiao2011}, indicate that fractional quantum hall states exist in the ground states of interacting lattice models, in the absence of an external magnetic field. It is found that the ground state is likely to respect the full lattice symmetry. Here we call these fractional ground states spin-polarized ``fractional Chern insulators'' (FCI) to distinguish from the traditional fractional quantum Hall states in an external magnetic field. These proposed lattice models share a common feature: a partially filled nearly flat two-dimensional band with non-trivial band topology.

The concept of band topology originates from the well-known TKNN index (or Chern number) of an IQH insulator\cite{Thouless1982}. In the past few years, this concept has been generalized to time-reversal symmetric systems, and triggers the theoretical and experimental discoveries of topological insulators in spin-orbital coupled compounds in both two and three spatial dimensions\cite{Qi2010,Hasan2010,Moore2010}. In two dimension (2D), a time-reversal symmetric band insulator is characterized by a $Z_2$ topological index. Experimentally, HgTe quantum heterostructure has been shown to be a 2D topological insulator\cite{Konig2007}. In the simplest limit, 2D topological insulator can be viewed as a direct product of the up-spin and down-spin wavefunctions hosting opposite TKNN index.

It is then quite natural to ask whether similar time-reversal-invariant (TRI) versions of two-dimensional fractional topological insulators (FTI) exist or not, and there has been a lot of interest in this issue\cite{Bernevig2006,Levin2009,Maciejko2010,Vaezi2011,Swingle2011}. In the simplest limit when spin along the $z$-direction is conserved, it can be understood as the direct product of wavefunctions of the up spin and the down spin with opposite FQHE. Clearly this direct product is a fully gapped stable phase. In addition it must have non-trivial ground state degeneracy on a torus even in the presence of a small $S_z$ conservation breaking perturbation, because the ground state degeneracy cannot be lifted by an arbitrary local perturbation. So there is no question that in principle this fractionalized phase could exist. One important issue is whether this phase hosts stable gapless edge excitations. This problem has been studied in by Levin and Stern\cite{Levin2009}. Another important open question is that whether TRI FTI can exist in a reasonable Hamiltonian.

In order to realize the FCI or FTI phases in experiments, one should find a compound with a nearly flat topological non-trivial band so that correlation effect is strong. Naively this is unnatural because usually a flat band is realized by spatially localized orbitals which do not support topological non-trivial hopping terms. However, a very recent theoretical investigation\cite{Xiao2011} on transition metal oxide heterostructures indicates that a nearly flat topological non-trivial band can be naturally realized in the $e_g$ orbital double-layer perovskite grown along the [111] direction. Exact diagonalization in the same work shows that fractional quantum hall state can be realized in principle when the nearly flat band is partially filled. Because the temperature scale of the FCI/FTI physics in this system is controlled by short-range Coulomb interaction, it can be a high-temperature effect.

Fractional quantum Hall states, especially the non-Abelian ones, have been shown to be very useful building blocks of quantum computers. If high temperature FCI/FTI physics can be realized experimentally, it will certainly have deep impact in condensed matter physics, including the efforts on topological quantum computation\cite{Nayak2008}.

Motivated by the recent progresses on FCI/FTI physics, in this paper we try to address several important issues: \emph{what are the many-particle wavefunctions of 2-D FCIs/FTIs? Can there be more than one FCI/FTI phases with the same filling fraction? If the answer is positive, can we classify these quantum phases (or ground state wavefunctions)?}

Historically, Laughlin's wavefunctions of FQH states in a magnetic field\cite{Laughlin1983} have been shown to be one of the most important theoretical progresses in many-particle physics. It allows people to understand a lot of properties of FQH liquids in a compact fashion, including the fractionalized quasi-particle excitations\cite{Arovas1984}, topological ground state degeneracies\cite{Haldane1985a,Wen1990b}, as well as constructing the low energy effective theories\cite{Girvin1987,Zhang1989,Read1989}. Here in the case of FCI/FTI systems, analytical understanding of the ground state wavefunctions will help us extract various measurable information in a similar way.

Recently there was an interesting work to construct FCI wavefunctions by proposing a one-to-one mapping between the lattice problem and the magnetic field problem\cite{Qi2011}. We would like to emphasize that the wavefunction problem for FCI is related to that for the magnetic field case, yet they are very different from each other. This is because the lattice symmetry of FCI is fundamentally different from the continuum case of the 2-D electron gas. In fact, the recently discovered FCI states preserve all the lattice point group symmetry as well as translational symmetry. \footnote{Because the wavefunction constructed in Ref.\cite{Qi2011} is based on one-dimensional Wannier function which explicitly select a special direction of the lattice, the constructed wavefunctions do not obviously respect the lattice point group symmetry.} Here in this paper, we point out that as a consequence of the lattice symmetry, there exist many different quantum FCI phases, all respecting the full lattice symmetry, even at the same filling fraction with the same quantum Hall conductance. These different FCI phases are distinct in the bulk in a more subtle way. One hand-waving statement is that the bulk quasi-particle excitations of these phases carry different lattice quantum numbers. These distinct FCI phases cannot be adiabatically connected with each other without a phase transition while the lattice symmetry is respected. Similar phenomena of distinct topologically ordered phases protected by symmetry is known in the context of quantum spin liquids\cite{Wen2002} and other low dimensional topological phases\cite{Chen2011}.

Now we outline the content of this paper. We start with the spin-polarized FCI at filling $\nu=\frac{1}{m}$ ($m$ is an odd number). In section \ref{PARTON} the $SU(m)$ parton construction of the fractional quantum Hall states (or spin-polarized FCI states) is introduced on a lattice, which is a natural generalization of the continuum case\cite{Laughlin1983,Jain1989a}. We argue that a general FCI wavefunction could break the $SU(m)$ gauge group down to $Z_m$, and consequently the low-energy dynamics is described by Chern-Simons-Higgs theories. We explicitly write down the form of the electronic FCI wavefunctions which will be useful for future variational Monte Carlo study. We construct quasiparticle excitations of such FCI states. To demonstrate how lattice symmetry restricts the structure of the wavefunctions, we introduce the concept of projective symmetry group (PSG)\cite{Wen2002} which serves as the mathematical language to classify different symmetry protected FCI phases.

With these theoretical preparations, in section \ref{EXAMPLE} we discuss one particular example, \ie the checkerboard lattice model\cite{Sun2011,Neupert2011} and write down two $SU(m)$ FCI wavefunctions and two $Z_m$ FCI wavefunctions in distinct universality classes for $\nu=1/3$. These wavefunctions support the same $\sigma_{xy}=\frac{1}{3}\frac{e^2}{h}$ quantized Hall conductance and similar topological properties. They are characterized by different PSGs in the bulk. These states can all serve as candidate states for the FCI state found in numerical simulations\cite{Neupert2011,Sheng2011,Regnault2011}. Which state is realized in the simulated model\cite{Neupert2011,Sheng2011,Regnault2011} would be determined by energetics. Because our proposed wavefunctions has the form of a Slater determinant and can be effectively implemented by variational Monte Carlo approach, the energetics of the proposed states can be studied by future numerical investigation. In Appendix \ref{app:example:honeycomb} we present another four examples of distinct FCI phases in the honeycomb lattice model\cite{Haldane1988}: two are $SU(m)$ states and the other two are $Z_m$ states.
We also propose spin-polarized FCI states with non-Abelian quasiparticles, which might be realized in nearly flat bands with Chern number $C>1$. Such non-Abelian FCIs might be used to build a universal quantum computer\cite{Freedman2002a,Nayak2008}.

In section \ref{EFT} we demonstrate that our parton construction can be used to compute the topological ground state degeneracy. This is particularly important for the $Z_m$ states, which belong to a new class of FQH wavefunctions.

In section \ref{FTI} we generalize our efforts to construct ground state wavefunctions of TRI FTIs. When the mixing between the up and down spins is weak in the electronic hamiltonian, it is natural to generalize our spin polarized results to this case. For filling fraction $\nu=\frac{2}{m}$ (on average $\nu=\frac{1}{m}$ for each spin), we present classes of $SU(m)^{\uparrow}\times SU(m)^{\downarrow}$ and $Z_m^{\uparrow}\times Z_m^{\downarrow}$ wavefunctions and discuss their properties including quasi-particle statistics and ground state degeneracies. We also propose a new parton construction formalism which allows one to write down generic electron wavefunctions for TRI FTI states in the absence of spin conservation. We can deform such a generic TRI FTI wavefunction in the absence of spin conservation into a $S^z$-conserved TRI FTI wavefunction (where spin-$\uparrow$ and spin-$\downarrow$ decouple) by continuously tuning a parameter. Stability of such a state against perturbations are briefly discussed.

\section{$SU(m)$ parton construction of spin-polarized fractional Chern insulator states}\label{PARTON}

\subsection{A brief review of Laughlin's FQH state from $SU(m)$ parton construction}

Soon after the experimental discovery of fractional quantum Hall (FQH) effects\cite{Tsui1982}, Laughlin proposed a series of variational wavefunctions\cite{Laughlin1983} which were shown\cite{Haldane1985} numerically to be a very good description of FQH states at odd-denominator filling fraction $\nu=1/m$. Later this idea of constructing trial wavefunctions was generalized to other filling fractions\cite{Haldane1983,Halperin1984,Jain1989}. An important lesson we can learn from Laughlin's wavefunction is as follows. With a fixed filling fraction (or a fixed number of flux quanta through the sample), the many-body wavefunction tends to vanish as fast as possible when two electrons approach each other so that the repulsive Coulomb energy between electrons could be minimized. As an example, Laughlin's state at $\nu=1/3$ is nothing but the cube of the wavefunction for a filled lowest Landau level. We can construct this wavefunction by splitting an electron into three fermionic \emph{partons}:
\begin{eqnarray}
&\label{electron:1/3}c({\bf r})=f_1({\bf r})f_2({\bf r})f_3({\bf r}).
\end{eqnarray}
Naturally from (\ref{electron:1/3}) we can see each parton carries $U(1)$ electric charge $e_0=e/3$ where $e$ stands for the electron charge.
The electron wavefunction is obtained through the following projection
\begin{eqnarray}\label{projection:1/3}
\Phi_e(\{{\bf r}_i\})=\langle0|\prod_{i=1}^Nf_1({\bf r}_i)f_2({\bf r}_i)f_3({\bf r}_i)|MF\rangle
\end{eqnarray}
where $|0\rangle$ represents the parton vacuum and $|MF\rangle$ can be any mean-field state of the three partons $f_{1,2,3}$. When each of the three partons occupy the lowest Landau level (LLL) one immediately obtains the Laughlin's state $\Phi_{\nu=1/3}(\{{\bf r}_i\})=\big[\Phi_{\nu=1}^\prime(\{{\bf r}_i\})\big]^3=\prod_{i<j}(z_i-z_j)^3\exp[-\sum_{i=1}^{N}|z_i|^2/(4l_B^2)]$. We have chosen the disc geometry and the symmetric gauge. $z_i=x_i+\imth y_i$ are complex coordinates, $l_B=\sqrt{\hbar/|eB|}=l_B^\prime/\sqrt3$ is the electron magnetic length and $l_B^\prime$ is the parton magnetic length.

Since each kind of parton occupies a LLL, the electro-magnetic response of the FQH state $\Phi_{\nu=1/3}(\{{\bf r}_i\})=\big[\Phi_{\nu=1}^\prime(\{{\bf r}_i\})\big]^3$ is characterized by Hall conductivity $\sigma_{xy}=3\cdot(\frac{e}3)^2/h=\frac13\frac{e^2}h$. This reproduces the correct filling fraction and the many-body Chern number. Note that electron operator $c({\bf r})$ in (\ref{electron:1/3}) is invariant under any \emph{local} $SU(3)$ transformation on the three partons $(f_1,f_2,f_3)^T$. The mean-field Hamiltonian density describing the Laughlin state is\cite{Wen1999}
\begin{eqnarray}
\mathcal{H}_{MF}=\frac1{2m^\ast}\sum_{\alpha=1}^3f_\alpha^\dagger({\bf r})\big(-\imth\nabla-e_0{\bf A}({\bf r})\big)^2f_\alpha({\bf r})\label{fqh:su3}
\end{eqnarray}
where $m^\ast$ is the effective mass of each parton. This mean-field Hamiltonian preserves the $SU(3)$ gauge symmetry and partons will also couple to a $SU(3)$ internal gauge field. Its effective theory is the $SU(3)_1$ Chern-Simons gauge theory, which explains the 3-fold topological ground state degeneracy on a torus\cite{Wen1998}. These $f^\dagger$ partons are nothing but charge $e/3$ quasiparticle excitations\cite{Laughlin1983} of Laughlin state. Indeed after projection (\ref{projection:1/3}) the three species of partons $f_{1,2,3}$ becomes indistinguishable thanks to the internal $SU(3)$ symmetry: each $f$ parton creates a charge $-e/3$ quasihole upon acting on the ground state $|MF\rangle$. It is straightforward to verify that the following wavefunction
\begin{eqnarray}\label{projection:1/3hole}
&\Phi_e(\{{\bf r}_i\}|{\bf w}_{1,2,3})=\\
&\langle0|f_1({\bf w}_1)f_2({\bf w}_2)f_3({\bf w}_3)\prod_{i=1}^{N-1}f_1({\bf r}_i)f_2({\bf r}_i)f_3({\bf r}_i)|MF\rangle\notag
\end{eqnarray}
reproduces the Laughlin wavefunction with three quasiholes at ${\bf w}_{1,2,3}$ up to a constant factor. Hence these partons are indeed charge $e/3$ anyons obeying fractional statistics with statistical angle $\theta_{1/3}=\frac\pi3$.

\subsection{$Z_m$ FCI state and its quasiparticles from $SU(m)$ parton construction}

Since the three seemingly-different partons $f_{1,2,3}$ are essentially the same quasihole excitations with the same quantum numbers, physically it is attempting to include the tunneling terms $f^\dagger_\alpha f_\beta,~\alpha\neq\beta$ in the mean-field Hamiltonian. By mixing different partons, these terms will break the internal $SU(3)$ gauge symmetry down to a a subgroup of $SU(3)$, which is $Z_3$, the center of the $SU(3)$ group, in the most generic case where $f^\dagger_\alpha f_\beta,~\forall~\alpha\neq\beta$ terms are present. In general the projected $Z_3$ wavefunction (\ref{projection:1/3}) is different from its parent projected $SU(3)$ wavefunction. For a 2-D electron gas in a magnetic field, however, people usually focus on the LLL within which the many-body wavefunction is an analytic function (\eg in the symmetric gauge on a disc). It is straightforward to show that as long as the mixing terms act inside the Hilbert space of LLL, the corresponding electron wavefunction (\ref{projection:1/3}) for a $Z_3$ state remains the same as that of its parent $SU(3)$ state. This is because the parton wavefunction describes a state with LLL fully filled. Mixing between different partons within the LLL Hilbert space only gives a unitary transformation of basis and does not modify the parton wavefunction. For a lattice model, it is natural to consider mixing terms acting between all bands (rather than within the filled bands), and the corresponding electron wavefunction of a $Z_3$ state will be a different wavefunction from that of its parent $SU(3)$ state. For a filling fraction $\nu=1/m$, our discussion straightforwardly generalizes to the corresponding $Z_m$ (the center of the $SU(m)$ group) state and its parent $SU(m)$ state.

To our knowledge, the $Z_m$ parton states of FQHE have not been proposed before. For this new class of wavefunctions, several natural questions need to be answered. What are the quasi-particles in the $Z_m$ state? What is the low-energy effective theory of the $Z_m$ state? Will it preserves the topological properties, such as ground state degeneracy? We answer these questions in this paper and find the topological properties of the $Z_m$ states are identical to the $SU(m)$ states. Their difference lies in the projective symmetry group, which is protected by lattice symmetry. In general, $Z_m$ states and $SU(m)$ states both serve as candidate ground states for the FCI states of a $\nu=1/m$ filled band with Chern number one.

To begin with, let us consider the quasiparticle excitations in a $Z_m$ state.
The physical quasiparticle excitations in a $Z_m$ state are constructed by inserting fluxes in the mean-field ansatz of $f^\dagger_\alpha f_\alpha$ terms and simultaneously creating vortices (or defects) in the Higgs condensates $\langle f^\dagger_\alpha f_\beta\rangle,~\alpha\neq\beta$.
In 2-D, because $\pi_1(SU(m)/Z_m)=Z_m$, these defects are the point-like vortices carrying $Z_m$ gauge fluxes.  Because the $Z_m$ flux can be considered to be localized in a single plaquette, one can effectively interpret it as a overall $U(1)$ gauge flux of all the $f$-partons. Namely when a $f$-parton winds around a fundamental vortex, it experiences a ${2\pi/m}$ flux. Due to the Chern numbers of the filled parton bands, this vortex also binds with a single $f$-parton gauge charge and thus carries electric charge $e/m$. Because this object carries both flux and gauge charge, the fractional statistical angle $\theta=\pi/m$ results. These are exactly the same electric charge and statistics that a quasiparticle carries in the $SU(m)$ state. We conclude that the topological properties of quasiparticles are the same in both the $Z_m$ and its $SU(m)$ parent state.

\begin{figure}
 \includegraphics[width=0.45\textwidth]{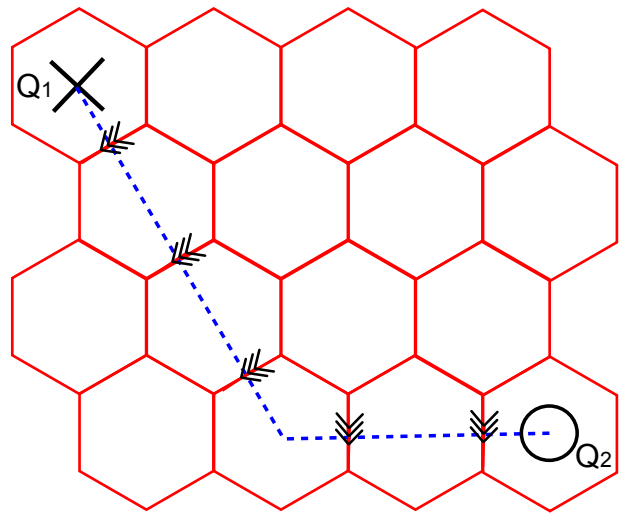}
\caption{(color online) A pair of anyonic quasiparticle and its antiparticle in a $Z_m$ state on the honeycomb lattice. The dashed line denotes the string of $e^{\imth2\pi k/m}$ phase shift connecting the two plaquettes where quasiparticle $Q_1$ and its antiparticle $Q_2$ are located. On top of the ground state mean-field ansatz, any mean-field bond crossing the string should pick up a phase shift $e^{\imth2\pi k/m}$. Here we only demonstrate the phase shift of nearest-neighbor (NN) mean-field amplitudes by triple arrows in the figure.}\label{fig:qp}
\end{figure}

Following the above discussions, we can write down the wavefunctions with low-energy anyonic excitations in a $Z_m$ state. At filling fraction $\nu=1/m$, in order to create one quasiparticle $Q_1$ at ${\bf w}_1$ and its antiparticle $Q_2$ at ${\bf w}_2$, we need to insert $2\pi k/m$ flux in a plaquette $P_{{\bf w}_1}$ at position ${\bf w}_1$ and $-2\pi k/m$ flux in a plaquette $P_{{\bf w}_2}$ at position ${\bf w}_2$ with $k=1,\cdots,m$. $Q_1$ carries $2\pi k/m$ flux and $e k/m$ charge while $Q_2$ carries $-2\pi k/m$ flux and $-ek/m$ charge. Both $Q_1$ and $Q_2$ have statistical angle $\theta=k^2\pi/m$ and their mutual statistical angle is $\theta^\prime=-k^2\pi/m$. They are realized by creating $e^{\imth2\pi k/m}$ phase shift for all mean-field amplitudes on the string connecting two plaquettes $P_{{\bf w}_1}$ and $P_{{\bf w}_2}$, on top of the mean-field ansatz for the ground state. An example of such a pair of quasiparticle and its antiparticle in a $Z_m$ state on honeycomb lattice is schematically shown in FIG. \ref{fig:qp}. The corresponding electron wavefunction is obtained by the projection of this new mean-field ansatz to the electronic degrees of freedom. When $m=3$, this projection is given by Eq.(\ref{projection:1/3}).

The ground state degeneracy of a $Z_m$ FCI state at $\nu=1/m$ on a torus can also be understood once we know its quasiparticle statistics\cite{Wen1990b,Oshikawa2006}. Consider the following tunneling process $\mathcal{T}_1$: a pair of quasiparticle (with flux $2\pi/m$ and charge $e/m$) and its antiparticle (with flux $-2\pi/m$ and charge $-e/m$) are created and the quasiparticle is dragged around the non-contractible loop $X_1$ along $x_1$ direction on the torus before it is finally annihilated with it anti-particle. This tunneling process will leave a string of $e^{\imth2\pi/m}$ phase shift (as shown in FIG. \ref{fig:qp}) along this loop $X_1$, therefore has the same physical effects as adiabatically inserting a $2\pi/m$ flux in the non-contractible loop $X_2$ along $x_2$ direction on the torus. Note that when the quasiparticle-anti-quasiparticle pair carries flux $\pm2\pi k/m$ and charge $\pm ke/m$ the corresponding tunneling process is realized by $\mathcal{T}_1^k$. Similarly we can define a tunneling process $\mathcal{T}_2$ by dragging the fundamental quasiparticle around non-contractible loop $X_2$ once, which is physically equivalent to inserting $2\pi/m$ flux in non-contractible loop $X_1$. In the thermodynamic limit, the Hilbert space of degenerate ground states should be expanded by these tunneling processes\cite{Oshikawa2006}. The two tunneling operators satisfy the following ``magnetic algebra":
\begin{eqnarray}
\mathcal{T}_1\mathcal{T}_2=\mathcal{T}_2\mathcal{T}_1e^{2\pi/m}
\end{eqnarray}
This is straightforward to understand from the point of view of Aharonov-Bohm effect. Another way to understand it is because the tunneling process $\mathcal{T}_1^{-1}\mathcal{T}_2^{-1}\mathcal{T}_1\mathcal{T}_2$ can de continuously deformed into two linked loops\cite{Wen1990b} and corresponds to a phase of $2\theta_{1/m}$, where $\theta_{1/m}=\pi/m$ is the statistical angle of the fundamental quasiparticle. All degenerate ground states can be labeled by \eg eigenvalues of unitary operators $\mathcal{T}_1$ and $\mathcal{T}_2^m$ (since they commute with each other). In this basis $\mathcal{T}_2$ acts like a ladder operator and changes the eigenvalue of $\mathcal{T}_1$ by a phase $e^{\imth2\pi/m}$. In this way one can see the ground state degeneracy of a $Z_m$ state on torus is $m$-fold. We can easily generalize this discussion to a genius-$g$ Riemann surface with $g$ pairs of non-contractible loops and the corresponding ground state degeneracy is $m^g$-fold. This is consistent with the ground state degeneracy calculated from the low-energy effective theory as will be shown in section \ref{EFT} in a formal way.

Because the discussion on low-energy effective theory of the $Z_m$ state involves more technical details, we postpone it to Section \ref{EFT}, where we compute its ground state degeneracy. We'll show that the ground state degeneracy of a $Z_m$ state is the same as that of a $SU(m)$ state: $m^g$-fold on a genus-$g$ Riemann surface.

\subsection{Regarding lattice symmetries}

In the numerical simulations of FCI phases\cite{Sheng2011,Neupert2011}, only the $m$-fold topological degeneracy of $\nu=1/m$ FCI is observed on torus. This indicates that the FCI wavefunctions respect the full lattice symmetry, since otherwise there should be extra degeneracies due to lattice symmetry breaking. This motivates us to write down the fully symmetric FCI wavefunctions.

In the following we outline the general strategy to construct fully symmetric FCI states on a lattice in the parton approach. Here we focus on spin-polarized FCI states with filling fraction $\nu=1/m$ through the $SU(m)$ parton construction. The electron operator is given by
\begin{eqnarray}\label{electron:1/m}
c({\bf r})=\prod_{\alpha=1}^mf_\alpha({\bf r})
\end{eqnarray}
where ${\bf r}$ is the coordinate of a lattice site. For simplicity we assume there is only one orbital per lattice site. As mentioned earlier, this parton construction has a local $SU(m)$ symmetry since the electron operator $c({\bf r})$ is invariant under any local transformation $f_\alpha({\bf r})\rightarrow\sum_{\beta}G_{\alpha\beta}({\bf r})f_\beta({\bf r})$ where $G({\bf r})\in SU(m)$. A generic parton mean-field ansatz is written as
\begin{eqnarray}
H^{MF}=\sum_{{\bf r},{\bf r}^\prime}\sum_{\alpha\beta}f^\dagger_\alpha({\bf r})M_{\alpha\beta}({\bf r}|{\bf r}^\prime)f_\beta({\bf r}^\prime)\label{mfH:1/m}
\end{eqnarray}
where $M({\bf r}|{\bf r}^\prime)=M^\dagger({\bf r}^\prime|{\bf r})$ is a $m\times m$ matrix assuming there are one electron orbital per site. Under a local $SU(m)$ gauge transformation $\{G({\bf r})\}$ it transforms as $M({\bf r}|{\bf r}^\prime)\rightarrow G({\bf r})M({\bf r}|{\bf r}^\prime)G^\dagger({\bf r}^\prime)$. Again once we obtain a mean-field state $|MF\rangle$ with the right filling number from (\ref{mfH:1/m}), the corresponding electron wavefunction is obtained through
\begin{eqnarray}
\Phi_e(\{{\bf r}_i\})=\langle0|\prod_{i=1}^Nc({\bf r}_i)|MF\rangle=\langle0|\prod_{i=1}^N\prod_{\alpha=1}^mf_\alpha({\bf r}_i)|MF\rangle ,\label{projection:1/m}
\end{eqnarray}
whose explicit form is a Slater determinant as given later in (\ref{wavefunction:zm}).

Not all parton mean-field ansatzs correspond to $\nu=1/m$ FCI states.
Let's start from a $SU(m)$ mean-field state with $M_{\alpha\beta}({\bf r}|{\bf r}^\prime)=\delta_{\alpha,\beta}T({\bf r}|{\bf r}^\prime)$: $H^{MF}_\alpha=\sum_{{\bf r},{\bf r}^\prime}f^\dagger_\alpha({\bf r})T_{\alpha,\beta}({\bf r}|{\bf r}^\prime)f_{\alpha}({\bf r}^\prime)$ where each flavor of the parton has the same filling number as the electron. For $\nu=1/m$ FCI states in topological flat bands, the filling fraction is such that on average there is one electron (hence one parton with each flavor) per $m$ unit cells. If the mean-field ansatz (\ref{mfH:1/m}) has explicit lattice translation symmetry, however, the corresponding state with $\nu=1/m$ filling would most likely be a gapless metallic state\footnote{When there are parton mixing terms which breaks the gauge symmetry from $SU(m)$ down to $Z_m$, one can construct a gapped state with filling $\nu=1/m$ by just filling the lowest parton band, since there is on average one parton (including all flavors) in each unit cell. However, the Hall conductance of such a state is $\sigma_{xy}=\frac{C}{m^2}\frac{e^2}h$ where $C$ is the Chern number of lowest parton band. Unless this lowest parton band has Chern number $C=m$ (which is unlikely), this gapped state at filling $\nu=1/m$ will have a Hall conductance different from $\sigma_{xy}=e^2/{(mh)}$ and is not a good candidate for the FCI states realized in recent numerical studies.} since only a fraction ($1/m$) of the lowest band is filled. How to construct a gapped mean-field ansatz of FCI with filling fraction $1/m$?

The answer lies in the $SU(m)$ gauge structure of the parton construction (\ref{electron:1/m}). This gauge structure allows the physical (lattice) symmetry to be realized projectively in the parton mean-field ansatz, which gives rise to a symmetric electron wavefunction after projection. Briefly speaking, the mean-field state itself can explicitly break lattice symmetries (such as lattice translations) while the electron wavefunction after projection (\ref{projection:1/m}) remains fully symmetric.

By inserting \eg $2\pi/m$ flux in each original unit cell, one can enlarge the unit cell by $m$ times. Therefore the corresponding mean-field state with filling $\nu=1/m$ is a state filling the lowest $m$ bands of mean-field ansatz (\ref{mfH:1/m}). If each of the $m$ lowest bands have a Chern number $+1$, the mean-field state filling these $m$ bands would have total Chern number $+m$, and the corresponding Hall conductivity is
\begin{eqnarray}
\sigma_{xy}=m\cdot(\frac1m)^2\cdot\frac{e^2}h=\frac{1}{m}\frac{e^2}{h},
\end{eqnarray}
because each parton carries $U(1)$ electric charge $e/m$. This gives the correct electromagnetic response of a $\nu=1/m$ spin-polarized FCI state.

Here the mean-field Hamiltonian (\ref{mfH:1/m}) explicitly breaks lattice translation symmetry due to the unit cell enlargement, but as long as the translated mean-field ansatz can be transformed to the original ansatz by a local $SU(m)$ gauge rotation, the corresponding electron wavefunction (\ref{projection:1/m}) still respect the translation symmetry. This is because any two mean-field ansatzs differ by a local $SU(m)$ gauge rotation give exactly the same electron wavefunction after projection. Similarly, even the mean-field Hamiltonian breaks other lattice symmetries such as the point group symmetry, the electron wavefunction after projection still can be fully symmetric. The mean-field ansatz simply forms a projective representation of the symmetry group. The mathematical framework of constructing fully symmetric electron wavefunctions based on parton mean-field ansatzs is the projective symmetry group (PSG), which will be introduced shortly. The technique of enlarging the unit cell by $m$ times without physically breaking any lattice symmetry will be generalized to the case of time-reversal-invariant FTI states with filling fraction $\nu=2/m$ in section \ref{FTI}.

Following this strategy, we always require that the parton mean-field ansatz of $\nu=1/m$ FCI state breaks lattice translation symmetry explicitly and enlarges the unit cell by $m$ times, so that the resultant mean-field state is an insulator. The partons will fill the lowest $m$ bands of the mean-field spectrum and the corresponding electron state after projection would still be gapped. Now that the number of momentum points of each band in the (reduced) 1st Brillouin zone equals the electron number $N$,  we can see the electron wavefunction (\ref{projection:1/m}) is nothing but a Slater determinant
\begin{align}\label{wavefunction:zm}
&\Phi_e(\{{\bf r}_i\})_{Z_m}=\det \mathcal{W}_{mN\times mN},\\
&\notag \mathcal{W}_{(\alpha-1)N+i,(n-1)N+j}=\phi_{\tk_j^n}({\bf r}_i^\alpha),\\
&\notag\alpha,n=1,\cdots,m;~~~i,j=1,\cdots,N.
\end{align}
where $\phi_{\tk_j^n}({\bf r}_i^\alpha)$ represents the eigenvector of mean-field Hamiltonian (\ref{mfH:1/m}). To be specific, $\phi_{\tk_j^n}({\bf r}_i^\alpha)$ corresponds to the $f_\alpha$ parton component in momentum-$\tk_j$ single-particle eigenvector of the bottom-up $n$-th band. Here $\alpha,n=1,\cdots,m$ and $i,j=1,\cdots,N$ where $N$ is the total electron number at filling fraction $\nu=1/m$. Note that for a $SU(m)$ mean-field ansatz (\ref{mfH:su(m)}) in the absence of mixing terms, the lowest $m$ bands are all degenerate and we have $\phi_{\tk^n}({\bf r}^\alpha)=\phi_\tk({\bf r})\delta_{n,\alpha}$. The corresponding electron wavefunction (\ref{wavefunction:zm}) reduces to the product of $m$ copies of a Slater determinant:
\begin{eqnarray}
\label{wavefunction:su(m)}
&\Phi_e(\{{\bf r}_i\})_{SU(m)}=\Big\{\det\begin{bmatrix}\phi_{\tk_1}({\bf r}_1)\cdots\phi_{\tk_N}({\bf r}_1)\\
\phi_{\tk_1}({\bf r}_2)\cdots\phi_{\tk_N}({\bf r}_2)\\
\cdots\cdots\cdots\\
\phi_{\tk_1}({\bf r}_N)\cdots\phi_{\tk_N}({\bf r}_N)\end{bmatrix}\Big\}^m
\end{eqnarray}
where $\phi_{\tk_j}({\bf r}_i)$ is the momentum-$\tk_j$ single-particle eigenvector of parton mean-field Hamiltonian $H_\alpha^{MF}=\sum_{{\bf r},{\bf r}^\prime}f^\dagger_\alpha({\bf r})T({\bf r}|{\bf r}^\prime)f_\alpha({\bf r}^\prime)$ with $\forall \alpha=1,\cdots,m$. This is a lattice version of Laughlin's state in free space\cite{Laughlin1983}. However once we add lattice-symmetry-preserving parton mixing terms which breaks gauge symmetry from $SU(m)$ to $Z_m$, the electron wavefunction (\ref{wavefunction:zm}) of a $Z_m$ FCI state, as well as its projective symmetry group which will be introduced shortly, will immediately become different from its parent $SU(m)$ state (\ref{wavefunction:su(m)}). We emphasize again that only when the unit cell is enlarged by $m$ times, we will have the same number of momentum points $\tk_j$ as the electron number $N$. The mean-field amplitudes can be determined by variational Monte Carlo study of the energetics of electronic wavefunctions (\ref{wavefunction:zm}).

Considering flux insertion in order to enlarge the unit cell in the mean-field ansatz (\ref{mfH:1/m}), another question arises: can there be more than one way of inserting fluxes into plaquettes without breaking physical lattice symmetries? If yes, how to classify different mean-field ansatzs (\ref{mfH:1/m})? The answer of the 1st question is positive and to answer the 2nd question, we need to introduce a mathematical structure: projective symmetry group (PSG)\cite{Wen2002} in order to characterize different ``universality classes" of symmetric FCI states. PSG classifies different mean-field ansatz which forms a projective representation of the physical symmetry group. In the following we give a brief introduction of PSG.

Note that there is a many-to-one correspondence between parton mean-field states and physical electron states due to the above projection operation: any two parton mean-field states related to each other by a $SU(m)$ gauge transformation $\{G(\bf r)\}$ correspond to the same electron state. As a result, although the physical electron state preserves all lattice symmetry, its parton mean-field ansatz may or may not explicitly preserve these lattice symmetries. The physical lattice symmetries are realized projectively in the mean-field ansatz. More precisely, in a generic case the parton mean-field ansatz (\ref{mfH:1/m}) should be invariant under a combination of lattice symmetry operation $U$ and a corresponding gauge transformation $\{G_U({\bf r})\in SU(m)\}$:
\begin{eqnarray}\label{mfH:symmetry}
M(U({\bf r})|U({\bf r}^\prime))=G_U(U({\bf r}))M({\bf r}|{\bf r}^\prime)G^\dagger_U(U({\bf r}^\prime))
\end{eqnarray}
Different universality classes of parton mean-field ansatzs are characterized by different PSGs\cite{Wen2002}, \ie different $SU(m)$ gauge transformations $\{G_U\}$ associated with symmetry operations $U$:
\begin{eqnarray}
PSG=\{G_U({\bf r})U|U\in~\text{symmetry group}\}
\end{eqnarray}
The low-energy gauge fluctuation of a mean-field ansatz is controlled by its invariant gauge group\cite{Wen2002} (IGG)
\begin{eqnarray}
IGG=\{G_\bse\in SU(m)|G_\bse M({\bf r}|{\bf r}^\prime)G^\dagger_\bse=M({\bf r}|{\bf r}^\prime),\forall~{\bf r}, {\bf r}^\prime\}\notag
\end{eqnarray}
where $\bse$ represents the identity operator of the (lattice) symmetry group (SG). In other words, IGG is a subgroup of the internal gauge group (which is $SU(m)$ here) that keeps the mean-field ansatz (\ref{mfH:1/m}) invariant. Hereafter we would call a parton mean-field state with \eg $IGG=SU(m)$ state a $SU(m)$ state. We can see that the IGG of mean-field ansatz (\ref{mfH:1/m}) always contains the following $Z_m$ group as a subgroup:
\begin{eqnarray}
Z_m=\{e^{\imth\frac{2\pi a}{m}}\cdot I_{m\times m}|a=1,2,\cdots,m\}\label{eq:Zm}
\end{eqnarray}
where $I_{m\times m}$ is the $m\times m$ identity matrix. This $Z_m$ group is the center of the $SU(m)$ group. A mean-field ansatz $\{M({\bf r}|{\bf r}^\prime)\}$ with $IGG=Z_m$ is called a $Z_m$ state\footnote{In principle the IGG of a mean-field ansatz can be any subgroup of the internal gauge group, \ie $SU(m)$ here. For example when $m=3$ by adding $f^\dagger_1f_2$ terms to a $SU(3)$ mean-field ansatz, the corresponding IGG becomes $U(1)$. Here we focus on the cases with $IGG=SU(m)$ or $IGG=Z_m$.}. The low-energy theory of this $Z_m$ state will be described by fermionic partons $f_\alpha$ interacting with $Z_m$ gauge fields.

The classification of PSGs with $IGG=SU(m)$ (which we call $SU(m)$ PSGs in this paper) are easy to carry out. The only gauge invariant quantities of a $SU(m)$ ansatz is the gauge-invariant flux through each plaquette, which must belong to the center of the $SU(m)$ gauge group, namely the $Z_m$ group in Eq.\ref{eq:Zm}, because otherwise the $SU(m)$ gauge group would be broken and $IGG$ cannot be $SU(m)$. Two $SU(m)$ states have the same PSG if and only if they have the same $Z_m$ gauge flux in each given plaquette. Therefore distinct $SU(m)$ PSGs have different $Z_m$ gauge flux pattern and vice versa.

The classification of PSGs with $IGG=Z_m$ (which we call $Z_m$ PSGs in this paper) involves more technical details and we leave this analysis for the honeycomb lattice model\cite{Haldane1988} and the checkerboard lattice model\cite{Sun2011,Neupert2011} in Appendix \ref{app:psg:honeycomb} and \ref{app:psg:checkerboard}.

\begin{figure}
 \includegraphics[width=0.22\textwidth]{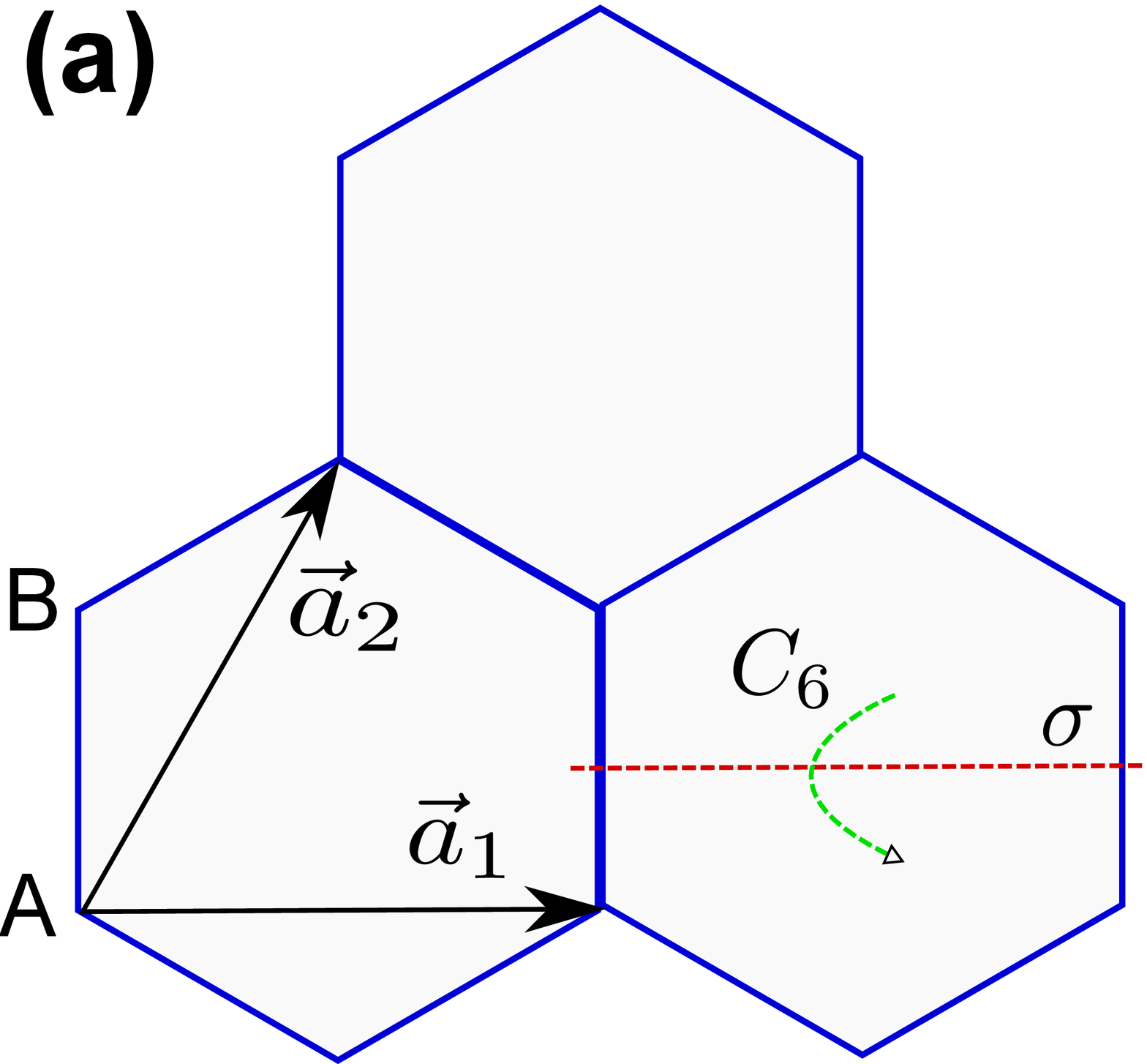}\;\includegraphics[width=0.22\textwidth]{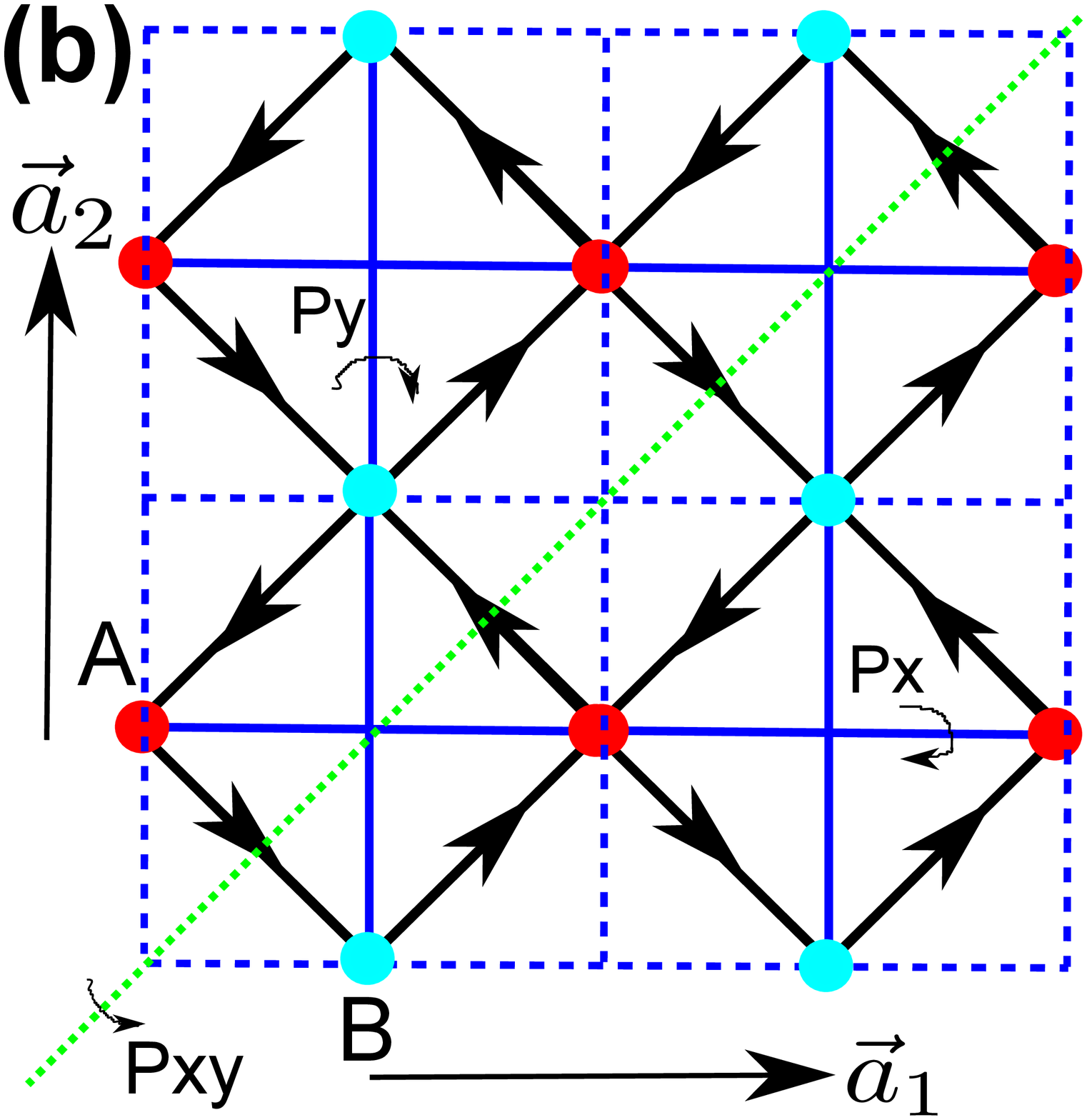}
\caption{(color online) The lattice structure and symmetry group generators of (a) honeycomb lattice and (b) checkerboard lattice. $\vec a_1$ and $\vec a_2$ are primitive lattice vectors. Each site is labeled by coordinate $(x,y,s)$: $\vec r=x\vec a_1+y\vec a_2$ corresponds to the position vector of its unit cell and $s=0/1$ are the sublattice indices for $A/B$ sublattices. $C_6$ represents $\pi/3$ rotations along $\hat z$-axis around the honeycomb plaquette center. $\bss$ and $P_{x},~P_y,~\pxy$ are $\pi$ rotations along the axis plotted in the figure.}\label{fig:lattice}
\end{figure}

\section{Examples}\label{EXAMPLE}

\begin{figure}
 \includegraphics[width=0.22\textwidth]{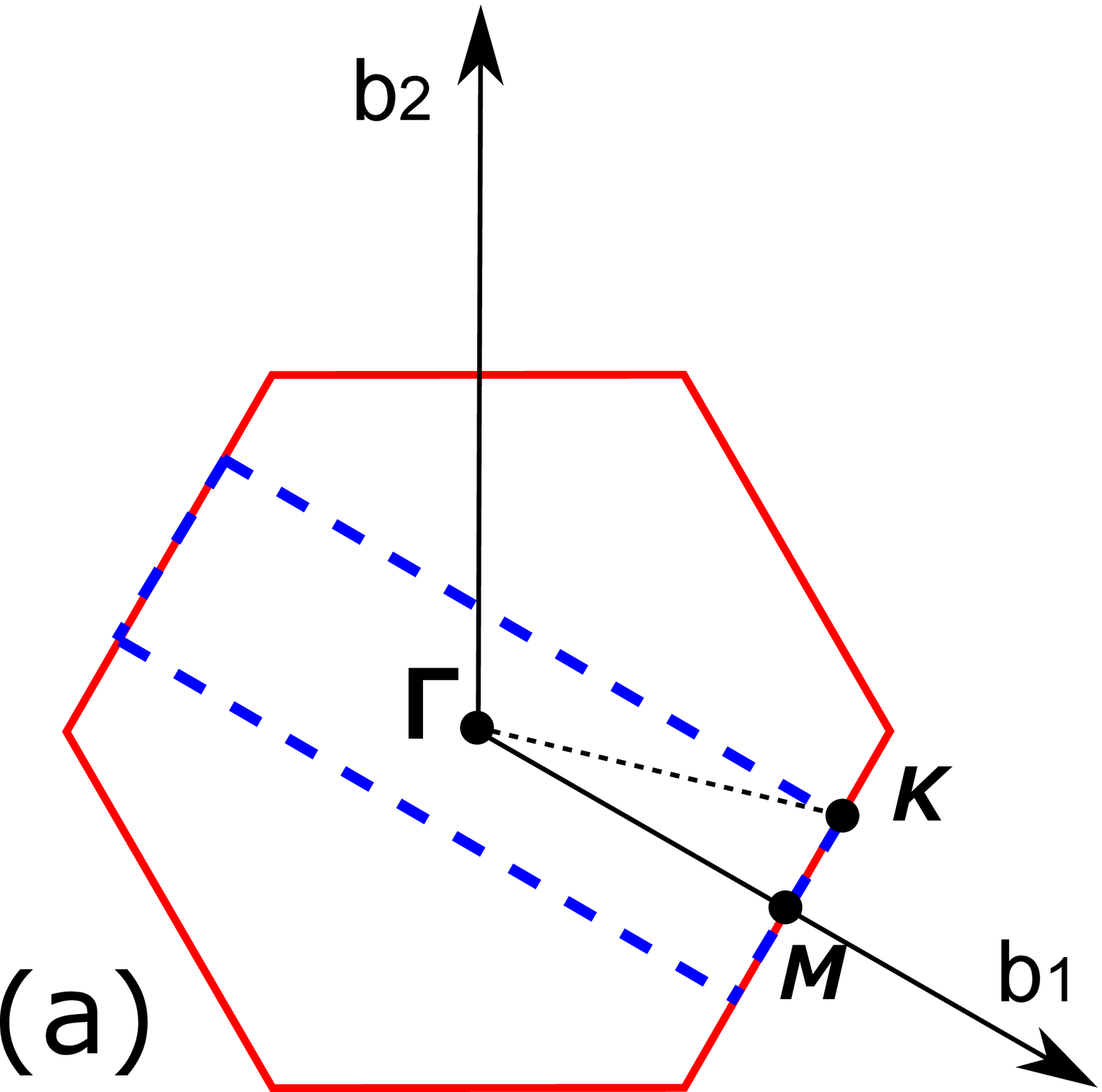}\;\includegraphics[width=0.2\textwidth]{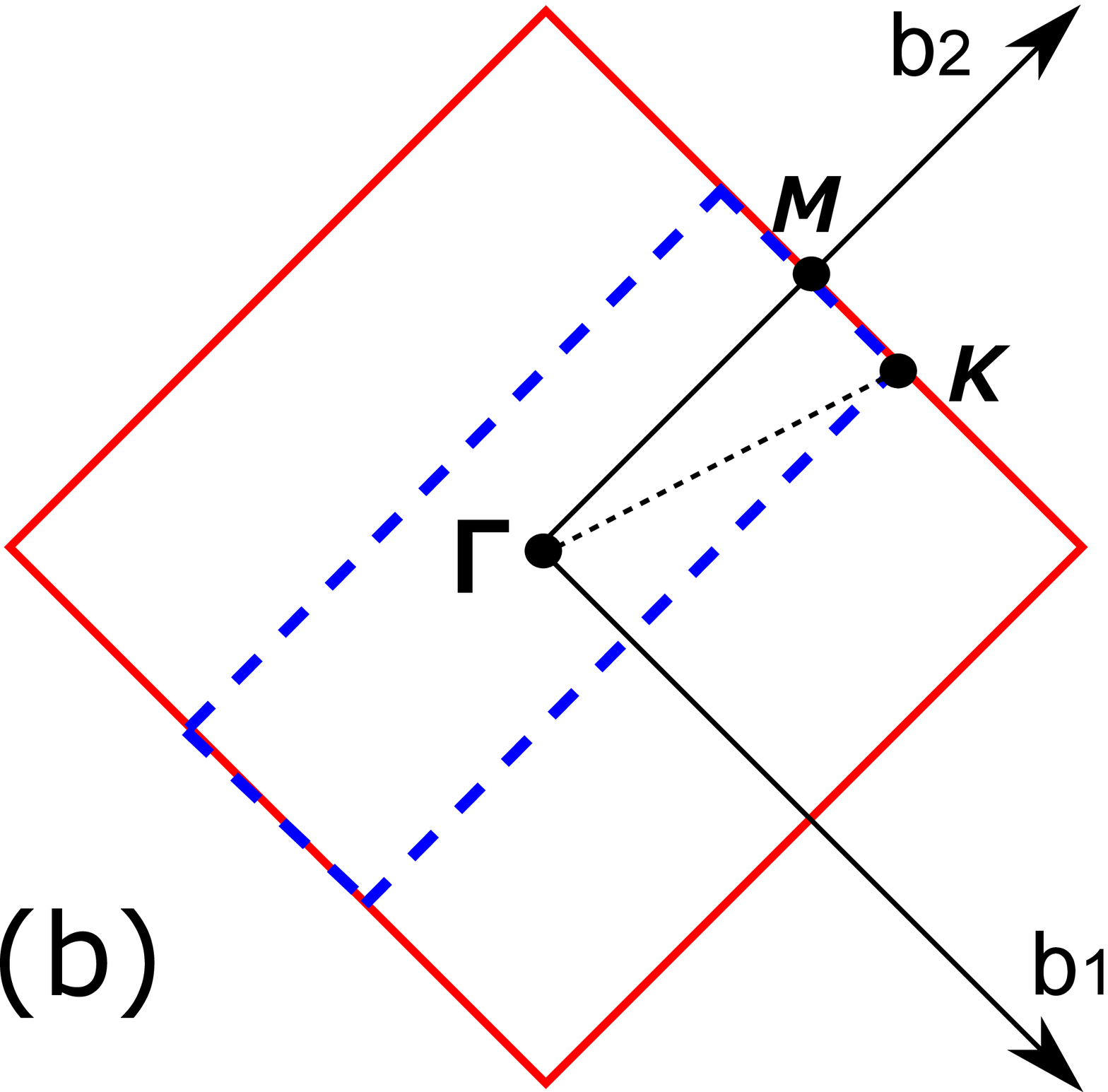}
\caption{(color online) The 1st Brillouin zone of (a) honeycomb lattice and (b) checkerboard lattice. $\vec b_1$ and $\vec b_2$ denoted by arrows are reciprocal lattice vectors satisfying $\vec b_i\cdot\vec a_j=\delta_{i,j}$ where $\vec a_{1,2}$ are primitive vectors shown in FIG. \ref{fig:lattice}. The solid line (red) denotes the original 1st Brillouin zone (BZ) of the two lattices while bold dashed line (blue) denotes the reduced 1st BZ when $\pm2\pi/3$ fluxes are inserted in each unit cell. The area of the reduced BZ is only one third of the original BZ. (a) For the honeycomb lattice, the three circles have momenta $\Gamma:~(k_1,k_2)=(0,0)$,~$K:~(k_1,k_2)=(7\pi/6,\pi/3)$ and $M:~(k_1,k_2)=(\pi,0)$, where $\tk=k_1\vec b_1+k_2\vec b_2$. (b) In the case of checkerboard lattice, the three circles have momenta $\Gamma:~(k_1,k_2)=(0,0)$,~$K:~(k_1,k_2)=(\pi/3,\pi)$ and $M:~(k_1,k_2)=(0,\pi)$, where $\tk=k_1\vec b_1+k_2\vec b_2$.}\label{fig:bz}
\end{figure}

The first part of this section shows four concrete examples of mean-field ansatzs corresponding to spin-polarized FCI states with filling fraction $\nu=1/3$ in the checkerboard lattice model\cite{Sun2011,Neupert2011}. These include two $Z_3$ ansatzs and their two $SU(3)$ parent states. As discussed in the previous section and proved in Appendix \ref{app:sym:honeycomb}-\ref{app:psg:checkerboard}, although the mean-field ansatz explicitly breaks lattice translation symmetry by tripling the unit cell, the physical electron state after projection (\ref{projection:1/m}) preserves all lattice symmetry. Four examples for the honeycomb lattice model with filling fraction $\nu=1/3$ are displayed in Appendix \ref{app:example:honeycomb}. We show the Hall conductance of all these states are $\sigma_{xy}=e^2/(3h)$.

In the second part of this section, we present a scenario which might realize non-Abelian FCI with $IGG=SU(m)$ by partially filling nearly flat bands with Chern number $C>1$, and two examples of such states in the $SU(3)$ parton construction are presented in Appendix \ref{app:example}.

\subsection{Four examples of spin-polarized FCI states with $\nu=1/3$ in the checkerboard lattice model}

The checkerboard lattice model\cite{Sun2011,Neupert2011} has been shown to support nearly flat bands with non-zero Chern numbers. Its symmetry group is shown in Appendix \ref{app:sym:checkerboard} and FIG. \ref{fig:lattice}. Each lattice site is labeled by coordinate $(x,y,s)$ as shown in Appendix \ref{app:sym:checkerboard}. Recently there are numerical evidence\cite{Neupert2011,Sheng2011,Regnault2011} of FCI states with filling fraction $\nu=1/3$ and $\nu=1/5$ in this model. By $SU(m)$ parton construction, we use PSG to classify different $Z_m$ mean-field ansatz ($m$ is an \emph{odd} integer) as shown in Appendix \ref{app:psg:checkerboard}. In the following we show two $Z_m$ mean-field states that belong to different universality classes. They correspond to two gauge-inequivalent solutions of (\ref{psg:checkerboard}) when $m=3$. We show mean-field amplitudes up to the next next nearest neighbor. Their two parent $SU(3)$ states also have different PSGs because they have different patterns of the $SU(3)$ gauge invariant fluxes. All four states are candidates of the FCI state found in the exact diagonalizations of checkerboard lattice model\cite{Neupert2011,Sheng2011,Regnault2011}.

\subsubsection{The $Z_3$ FCI state CB1 with $\sigma_{xy}=1/3$ and its parent $SU(3)$ state}

\begin{figure}
 \includegraphics[width=0.4\textwidth]{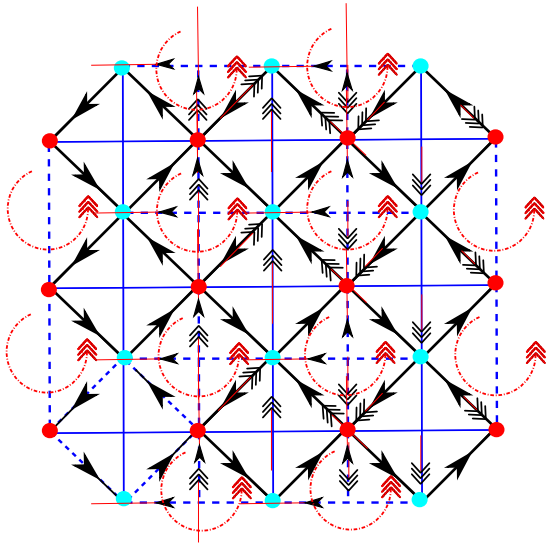}
\caption{(color online) Mean-field ansatz of the parent $SU(3)$ parton states associated with FCI state CB1 on checkerboard lattice. The solid lines on NN bonds represent complex hopping amplitude $\alpha$ along the direction of the arrow. Solid lines on NNN bonds represents real hopping amplitude $\beta_x$. The dashed lines on NNN bonds represent complex hopping amplitude $e^{\imth\pi/3}\beta_y$ along the direction of the arrow. The triple arrow means the original hopping amplitude along its direction should be multiplied by a phase factor $\eta_{12}$. We only show up to NNN terms and all NNNN terms can be obtained from (\ref{mfH:symmetry}). Note that in the mean-field ansatz the lattice translation along $\vec a_1$ direction is explicitly broken by flux insertion and the unit cell is tripled. However the lattice translation symmetry is preserved in the corresponding electron states after projection (\ref{projection:1/m}).}\label{fig:cb1}
\end{figure}

\begin{figure}
 \includegraphics[width=0.45\textwidth]{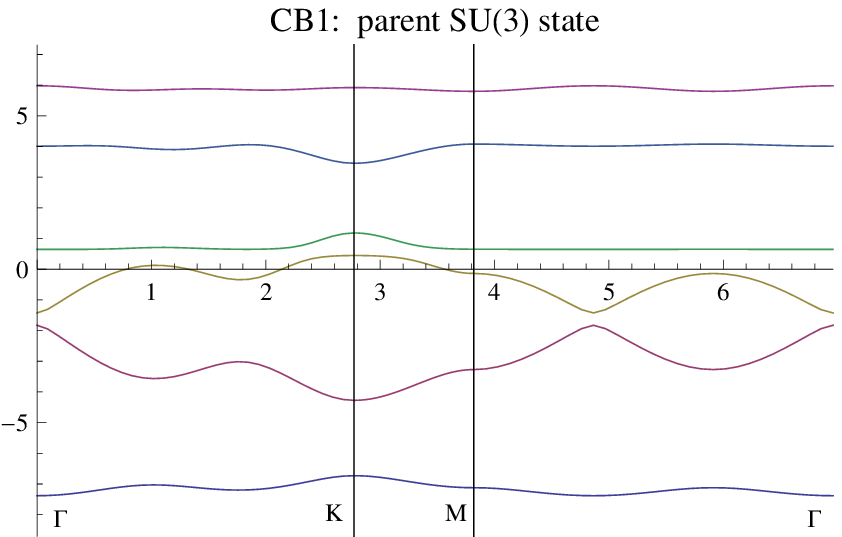}
\caption{(color online) The parton band structure of the parent $SU(3)$ ansatz (\ref{mfH:su(m)}) of CB1 state. Each band is 3-fold degenerate, corresponding to 3 parton flavors $f_{1,2,3}$. We plot the dispersion along $\Gamma\rightarrow K\rightarrow M\rightarrow\Gamma$ as shown in FIG. \ref{fig:bz}(b). Hopping parameters are chosen as $\alpha=e^{\imth\pi/12}$ for NN,~$\beta_x=-0.2=\beta_y$ for NNN and $\gamma=0.1$ for NNNN. The Chern numbers of the 6 bands are $\{1,-2,-2,4,-2,1\}$ in a bottom-up order.}\label{fig:cb1_band}
\end{figure}

In the $Z_3$ FCI state CB1 the gauge transformations $G_U(x,y,s)$ associated with lattice symmetries $U$ are listed below:
\begin{eqnarray}
&G_{T_2}(x,y,s)=I_{3\times3},~~~G_{T_1}(x,y,s)=\eta_{12}^yI_{3\times3},\\
&\notag G_{P_x}(x,y,s)=\eta_{12}^{sx}I_{3\times3},\\
&\notag G_{P_y}(x,y,s)=I_{3\times3},~~~G_{\pxy}(x,y,s)=\eta_{12}^{xy}I_{3\times3}.
\end{eqnarray}
where $\eta_{12}=\exp(-\imth2\pi/3)$. $T_{1,2}$ are lattice translation operations and $P_{x,y,xy}$ are reflections. They are schematically shown in Fig.\ref{fig:lattice}(b) and defined mathematically in Appendix \ref{app:sym:checkerboard}. As shown in Appendix \ref{app:psg:checkerboard} the symmetry allowed mean-field amplitudes are:

(\Rmnum{1}) For nearest neighbor (NN) amplitude $u_\alpha\equiv M(0,0,1|0,0,0)$
\begin{eqnarray}
u_\alpha=u_\alpha^T
\end{eqnarray}
\ie $u_\alpha$ can be any complex symmetric $3\times3$ matrix. All other NN amplitudes can be generated from $u_\alpha$ by symmetry operations through (\ref{mfH:symmetry}).

There are two independent NNN amplitudes:

(\Rmnum{2}a) For next nearest neighbor (NNN) amplitude $u_{\beta x}\equiv M(1,0,0|0,0,0)$
\begin{eqnarray}
u_{\beta x}=u_{\beta x}^T=u_{\beta x}^\ast
\end{eqnarray}
\ie $u_{\beta x}$ can be any real symmetric $3\times3$ matrix. Half of NNN mean-field amplitudes can be generated from $u_{\beta x}$ by symmetry operations through (\ref{mfH:symmetry}).

(\Rmnum{2}b) For next nearest neighbor (NNN) amplitude $u_{\beta y}\equiv M(0,1,0|0,0,0)$
\begin{eqnarray}
u_{\beta y}=e^{\imth\pi/3}\tilde u_{\beta y}
\end{eqnarray}
where $\tilde u_{\beta y}$ can be any real symmetric $3\times3$ matrix. Half of NNN mean-field amplitudes can be generated from $u_{\beta y}$ by symmetry operations through (\ref{mfH:symmetry}).

(\Rmnum{3}) For next next nearest neighbor (NNNN) amplitude $u_\beta\equiv M(1,1,0|0,0,0)$
\begin{eqnarray}
u_\gamma=u_\gamma^\dagger
\end{eqnarray}
\ie $u_\gamma$ can be any Hermitian $3\times3$ matrix. All other NNNN mean-field amplitudes can be generated from $u_\gamma$ by symmetry operations through (\ref{mfH:symmetry}).

These mean-field amplitudes of $u_\alpha$, $u_{\beta x}$, $u_{\beta y}$ and $u_\gamma$ should be treated as variational parameters. Their optimal values which minimize the variational energy of electron wavefunction (\ref{wavefunction:zm}) can be determined in variational Monte Carlo simulations.

\begin{figure}
 \includegraphics[width=0.45\textwidth]{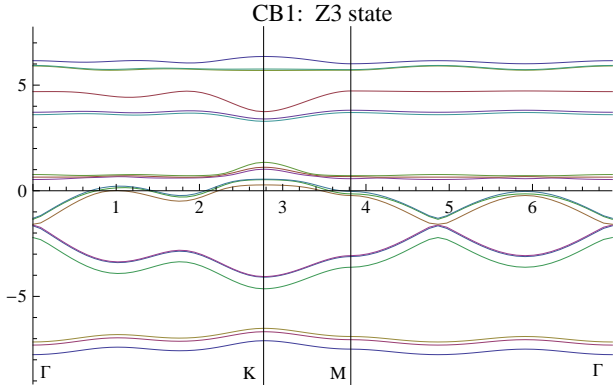}
\caption{(color online) The parton band structure of a $Z_3$ mean-field ansatz (\ref{mfH:1/m}): the CB1 state. We plot the dispersion along $\Gamma\rightarrow K\rightarrow M\rightarrow\Gamma$ as shown in FIG. \ref{fig:bz}(b). Note that each 3-fold degenerate band in its parent $SU(3)$ ansatz (see FIG. \ref{fig:cb1_band}) now splits into 3 non-degenerate parton bands. All the 3 lowest bands have Chern number $+1$.}\label{fig:cb1_z3}
\end{figure}

The corresponding parent $SU(3)$ state has $u_\alpha=\alpha I_{3\times3},~u_{\alpha x}=\beta_x I_{3\times3},~u_{\alpha y}=e^{\imth\pi/3}\beta_y I_{3\times3}$ and $u_{\gamma}=\gamma I_{3\times3}$. Choosing parameters $\alpha=e^{\imth\pi/12}$,~$\beta_x=-0.2=\beta_y$ and $\gamma=0.1$ we have the Chern numbers of the 6 bands for each parton species as $\{1,-2,-2,4,-2,1\}$ and the lowest band is well separated from other bands, as shown in FIG. \ref{fig:cb1_band}. This qualitative band structure persists for a large parameter range. Each band of the $SU(3)$ parton ansatz is 3-fold degenerate, corresponding to the 3 parton flavors $f_{1,2,3}$.  By adding small parton mixing terms to the $SU(3)$ mean-field state, each 3-fold degenerate band splits into 3 non-degenerate parton bands in a $Z_3$ mean-field state, as shown in FIG. \ref{fig:cb1_z3}. By filling the resulting 3 lowest bands (all with Chern number $+1$) we obtain a $Z_3$ FCI state whose Hall conductivity is $\sigma_{xy}=1/3$ in the unit of $e^2/h$.

\subsubsection{The $Z_3$ FCI state CB2 with $\sigma_{xy}=1/3$ and its parent $SU(3)$ state}

\begin{figure}
 \includegraphics[width=0.4\textwidth]{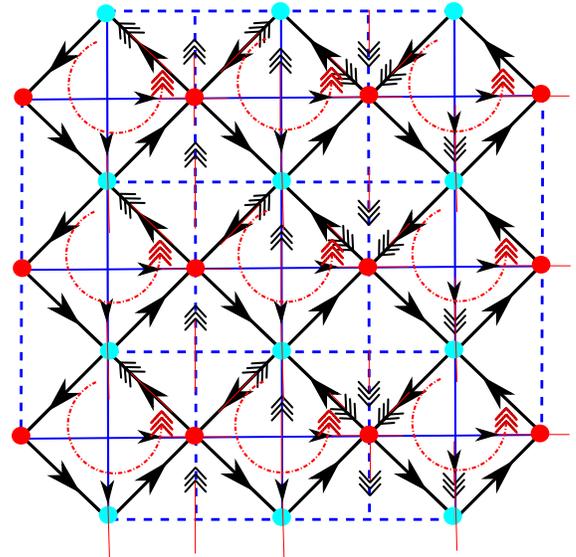}
\caption{(color online) Mean-field ansatz of the parent $SU(3)$ parton states associated with FCI states CB2 and CB3 on checkerboard lattice. The solid lines on NN bonds represent complex hopping amplitude $\alpha$ along the direction of the arrow. Dashed lines on NNN bonds represents real hopping amplitude $\beta_y$.
Solid lines on NNN bonds represent complex hopping amplitude $e^{\imth\pi/3}\beta_x$ along the direction of the arrow. The triple arrow means the original hopping amplitude along its direction should be multiplied by a phase factor $\eta_{12}$. We only show up to NNN terms and all NNNN terms can be obtained from (\ref{mfH:symmetry}). Note that in the mean-field ansatz the lattice translation along $\vec a_1$ direction is explicitly broken by flux insertion and the unit cell is tripled. However the lattice translation symmetry is preserved in the corresponding electron states after projection (\ref{projection:1/m}).}\label{fig:cb23}
\end{figure}

In the $Z_3$ FCI state CB2 the gauge transformations $G_U(x,y,s)$ associated with lattice symmetry $U$ are listed below:
\begin{eqnarray}
&G_{T_2}(x,y,s)=I_{3\times3},~~~G_{T_1}(x,y,s)=\eta_{12}^yI_{3\times3},\\
&\notag G_{P_x}(x,y,s)=\eta_{12}^{(s-1)x}I_{3\times3},\\
&\notag G_{P_y}(x,y,s)=\eta_{12}^yI_{3\times3},~~~G_{\pxy}(x,y,s)=\eta_{12}^{xy}I_{3\times3}.
\end{eqnarray}
where $\eta_{12}=\exp(-\imth2\pi/3)$. As shown in Appendix \ref{app:psg:checkerboard} the symmetry allowed mean-field amplitudes are:

(\Rmnum{1}) For nearest neighbor (NN) amplitude $u_\alpha\equiv M(0,0,1|0,0,0)$
\begin{eqnarray}
u_\alpha=u_\alpha^T
\end{eqnarray}
\ie $u_\alpha$ can be any complex symmetric $3\times3$ matrix. All other NN amplitudes can be generated from $u_\alpha$ by symmetry operations through (\ref{mfH:symmetry}).

There are two independent NNN amplitudes:

(\Rmnum{2}a) For next nearest neighbor (NNN) amplitude $u_{\beta x}\equiv M(1,0,0|0,0,0)$
\begin{eqnarray}
u_{\beta x}=e^{\imth\pi/3}\tilde u_{\beta x}
\end{eqnarray}
where $\tilde u_{\beta x}$ can be any real symmetric $3\times3$ matrix. Half of NNN mean-field amplitudes can be generated from $u_{\beta x}$ by symmetry operations through (\ref{mfH:symmetry}).

(\Rmnum{2}b) For next nearest neighbor (NNN) amplitude $u_{\beta y}\equiv M(0,1,0|0,0,0)$
\begin{eqnarray}
u_{\beta y}=u_{\beta y}^T=u_{\beta y}^\ast
\end{eqnarray}
\ie $u_{\beta y}$ can be any real symmetric $3\times3$ matrix. Half of NNN mean-field amplitudes can be generated from $u_{\beta y}$ by symmetry operations through (\ref{mfH:symmetry}).

(\Rmnum{3}) For next next nearest neighbor (NNNN) amplitude $u_\beta\equiv M(1,1,0|0,0,0)$
\begin{eqnarray}
u_\gamma=u_\gamma^\dagger
\end{eqnarray}
\ie $u_\gamma$ can be any Hermitian $3\times3$ matrix. All other NNNN mean-field amplitudes can be generated from $u_\gamma$ by symmetry operations through (\ref{mfH:symmetry}).

\begin{figure}
 \includegraphics[width=0.45\textwidth]{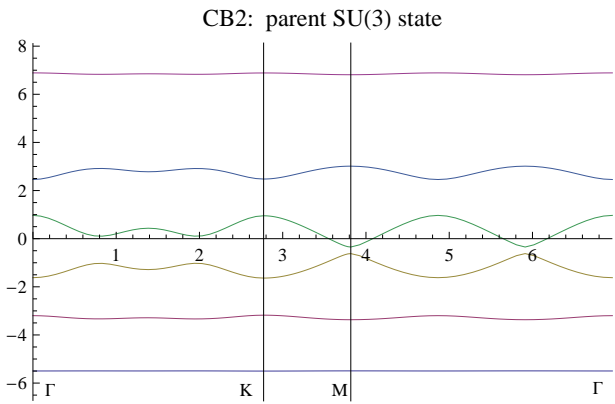}
\caption{(color online) The parton band structure of the parent $SU(3)$ ansatz (\ref{mfH:su(m)}) of CB2 state. Each band is 3-fold degenerate, corresponding to 3 parton flavors $f_{1,2,3}$. We plot the dispersion along $\Gamma\rightarrow K\rightarrow M\rightarrow\Gamma$ as shown in FIG. \ref{fig:bz}(b). Hopping parameters are chosen as $\alpha=e^{\imth\pi/12}$ for NN,~$\beta_x=0.1=\beta_y$ for NNN and $\gamma=0.09$ for NNNN. The Chern numbers of the 6 bands are $\{1,1,1,-5,1,1\}$ in a bottom-up order.}\label{fig:cb2_band}
\end{figure}

The corresponding parent $SU(3)$ state has $u_\alpha=\alpha I_{3\times3},~u_{\alpha x}=e^{\imth\pi/3}\beta_x I_{3\times3},~u_{\alpha y}=\beta_y I_{3\times3}$ and $u_{\gamma}=\gamma I_{3\times3}$. From FIG. \ref{fig:cb23} we can see that the pattern of fluxes in CB2 state is different from that in CB1 state. For example, in CB1 state $-2\pi/3$ flux (shown by triple arrows) are inserted in one half of the NN square plaquettes (those enclosed by two horizontal blue sites and two vertical red sites) while in CB2 state $-2\pi/3$ flux are inserted in the other half of the NN square plaquettes (those enclosed by two horizontal red sites and two vertical blue sites). Notice that these two types of NN square plaquettes are inequivalent, not related to each other by any lattice symmetry (such as translation or mirror reflection) in the checkerboard lattice model\cite{Sun2011}. Therefore CB1 and CB2 are different mean-field $Z_3$ FCI states belonging to distinct universality classes.  Choosing parameters $\alpha=e^{\imth\pi/12}$,~$\beta_x=0.1=\beta_y$ and $\gamma=0.09$ we have the Chern numbers of the 6 bands for each parton species as $\{1,1,1,-5,1,1\}$, and the lowest band is well separated from other bands, as illustrated in FIG. \ref{fig:cb2_band}. This qualitative band structure persists for a large parameter range. Each band of a $SU(3)$ parton ansatz is 3-fold degenerate, corresponding to the 3 parton flavors $f_{1,2,3}$. By adding small parton mixing terms to the $SU(3)$ mean-field state, each 3-fold degenerate band splits into 3 non-degenerate parton bands in a $Z_3$ mean-field state. By filling the resultant 3 lowest bands (all with Chern number $+1$) we obtain a $Z_3$ FCI state whose Hall conductivity is again $\sigma_{xy}=1/3$ in unit of $e^2/h$.

\subsection{Possible non-Abelian states by partially filling a nearly flat band with Chern number $C>1$}

It has been shown\cite{Wen1991b,Wen1998} that in $SU(m)$ parton construction, when each parton species fills $n$ Landau levels, the effective theory of the corresponding electron state is the $SU(m)_n$ Chern-Simons theory and the system has non-Abelian quasiparticle excitations when $n>1$. Moreover, the non-Abelian quasiparticles of this state can be used as topologically-protected qubits in a universal quantum computer\cite{Nayak2008} as long as $m>2$. This motivates us to propose possible realization of non-Abelian FCI states realized in a partially-filled nearly flat band with Chern number $C>1$. When this band is partially filled with a filling fraction \eg $\nu=1/m$, the Hall conductance of the corresponding FCI state would be $\sigma=\frac Cm\frac{e^2}h$. If the lowest $m$ parton bands all have Chern number $C$ instead of $+1$, the $SU(m)$ FCI state obtained by filling $m$ lowest parton bands indeed has Hall conductance $\sigma=m\frac{C e_0^2}h=\frac Cm\frac{e^2}h$ with $e_0=e/m$. This $SU(m)$ FCI state can be a promising candidate for $\nu=1/m$-filled nearly flat band with $C>1$.

We discuss the $C=2$ case as an example. In Appendix \ref{app:example} using the $SU(3)$ parton construction, we show two examples of $SU(3)$ FCIs, one on the honeycomb lattice and the other on the checkerboard lattice, both have 3 lowest parton bands with Chern number $+2$. These two $SU(3)$ FCIs with $\nu=1/3$ and $\sigma=\frac23\frac{e^2}h$ are non-Abelian FCIs. Their low-energy effective theory of these $SU(3)$ states is $SU(3)_2$ Chern-Simons theory\cite{Wen1998}, featured by $6$-fold ground state degeneracy on the torus and non-Abelian quasiparticle excitations\cite{Witten1989}. These results indicate that once a nearly flat band with Chern number $C>1$ is found, by partially filling it one may realize non-Abelian FCIs, which have the potential to build a universal quantum computer\cite{Freedman2002a,Nayak2008}.

\section{Effective theory and ground state degeneracy of spin-polarized $Z_m$ FCI states}\label{EFT}

As mentioned earlier, the partons in our construction not only couples to the external electromagnetic gauge field, but also couples to a dynamical internal gauge field, which is a $SU(m)$ ($Z_m$) gauge field for a $SU(m)$($Z_m$) state. The low energy effective theories of the partons coupled with the internal gauge fields control all the topological properties of the systems. The topological properties of a $SU(m)$ state has been studied before\cite{Wen1998,Wen1999}. In this section we will analyze the low-energy effective theory of spin-polarized $Z_m$ FCI state (\ref{mfH:1/m}) from the $SU(m)$ parton construction. To our knowledge, the $Z_m$ FQH states have not been proposed and studied before. We'll try to answer the following questions: what is the ground state degeneracy of the $Z_m$ FCI state? Is it the same as or different from that of a $SU(m)$ FCI state?


We start from a $SU(m)$ mean-field state which has been shown\cite{Wen1998} to describe the $\nu=1/m$ Laughlin state in the continuum limit. Its mean-field ansatz is
\begin{eqnarray}\label{mfH:su(m)}
M_{\alpha,\beta}({\bf r}|{\bf r}^\prime)=\delta_{\alpha,\beta}~T({\bf r}|{\bf r}^\prime)
\end{eqnarray}
In other words, there is no hopping between partons of different species and the $m$ species of partons have exactly the same band structure. As shown in (\ref{wavefunction:su(m)}) its electron wavefunction is a lattice version of Laughlin state\cite{Laughlin1983}. Apparently this mean-field state doesn't break the $SU(m)$ gauge symmetry which leaves the electron operator (\ref{electron:1/m}) invariant. Since here the partons couple to both the $U(1)$ electromagnetic field and the $SU(m)$ gauge field, the Lagrangian writes
\begin{eqnarray}
&L_{SU(m)}=\\
&\notag\int\text{d}t\Big\{\sum_{\alpha}\sum_{\bf r}\imth f^\dagger_\alpha({\bf r},t)\partial_t f_\alpha({\bf r},t)-\sum_{\alpha,{\bf r},{\bf r}^\prime}f^\dagger_\alpha({\bf r},t)T({\bf r}|{\bf r}^\prime)\\
&\notag\cdot e^{-\imth e_0\int_{\bf r}^{{\bf r}^\prime}\vec A({\bf x},t)\cdot\vec{\text{d}{\bf x}}}\mathcal{P}\big[e^{-\imth\int_{\bf r}^{{\bf r}^\prime}\vec a({\bf x},t)\cdot\vec{\text{d}{\bf x}}}\big]f_\alpha({\bf r}^\prime,t)\Big\},
\end{eqnarray}
where $\mathcal{P}$ means path-ordered integral. $A_\mu$ and $a_\mu$ are the $U(1)$ and the $SU(m)$ gauge fields respectively. Strictly speaking they are both defined on the link of a lattice here. To linear order the above action can be written as
\begin{eqnarray}
&L_{SU(m)}=\\
&\int\text{d}t\Big\{\sum_{\alpha}\sum_{\bf r}f^\dagger_\alpha({\bf r},t)\Big[\imth\partial_t f_\alpha({\bf r},t)-\sum_{{\bf r}^\prime}T({\bf r}|{\bf r}^\prime) f_\alpha({\bf r}^\prime,t)\Big]\notag\\
&\notag-e_0\sum_{\bf r}J^{U(1)}_\mu({\bf r},t)A^\mu({\bf r},t)\\
&\notag-\sum_{\bf r}\big(J^{SU(m)}_\mu\big)_{\alpha,\beta}({\bf r},t)~a^\mu_{\alpha,\beta}({\bf r},t)\Big\}
\end{eqnarray}
where $A^\mu({\bf r})$ stands for electromagnetic $U(1)$ gauge field while $a^\mu({\bf r})$ represents the internal $SU(m)$ gauge field. $e_0=e/m$ is the electric charge of each parton. Here $J^{U(1)}({\bf r})$ and $J^{SU(m)}_\mu({\bf r})$ are conserved $U(1)$ and $SU(m)$ parton currents respectively. To be precise, $J_0^{U(1)}=\sum_\alpha f^\dagger_\alpha f_\alpha$ and $(J_0^{SU(m)})_{\alpha\beta}=f^\dagger_\alpha f_\beta-\delta_{\alpha,\beta}\sum_{\gamma}f^\dagger_\gamma f_\gamma/m$. In the long-wavelength limit the spatial components of the parton currents in momentum space (with momentum $\tq$) writes:
\begin{eqnarray}
&\notag \overrightarrow{J^{U(1)}_{\mu,\tq}}=
\sum_{\tk}\vec \nabla_\tk T_\tk\sum_\alpha f^\dagger_{\alpha,\tk-\tq/2}f_{\alpha,\tk+\tq/2},\\
&\notag \big(\overrightarrow{J^{SU(m)}_{\mu,\tq}})_{\alpha,\beta}=
\sum_{\tk}\vec\nabla_\tk T_\tk f^\dagger_{\alpha,\tk-\tq/2}f_{\beta,\tk+\tq/2}-\delta_{\alpha,\beta}\frac{\overrightarrow{J^{U(1)}_{\mu,\tq}}}m.
\end{eqnarray}
Since the partons form a band insulator, the band gap allows us to safely integrate out the partons and obtain an effective action $L[A,a]$ for the gauge fields. Let's assume all the filled $m$ lowest parton bands have Chern number $+1$. Upon integrating out partons $\{f_\alpha,f^\dagger_\alpha\}$, the effective Lagrangian density writes
\begin{eqnarray}
&\mathcal{L}_{SU(m)}[A,a]=\frac{me_0^2}{4\pi}\epsilon_{\mu\nu\lambda}A_\mu\partial_\nu A_\lambda\\
&\notag+\frac1{4\pi}\epsilon_{\mu\nu\lambda}\text{Tr}\big(a_\mu\partial_\nu a_\lambda+\frac\imth3a_\mu a_\nu a_\lambda\big)
\end{eqnarray}
the first term corresponds to the quantized Hall conductance $\sigma_{xy}=m e_0^2/h$, while the second term, \ie a $SU(m)_1$ Chern-Simons term describes the low-energy gauge fluctuations. As shown in Appendix \ref{app:g.s.deg}, the Chern-Simons theory of $SU(m)$ gauge field $a_\mu$ can be reduced to Chern-Simons theory of $U(1)$ gauge fields $a_\mu^I,~I=1,\cdots,m-1$. The gauge field configuration is given by $(a_\mu)_{\alpha,\beta}=\sum_{I=1}^{m-1}a^I_\mu g^I_{\alpha,\beta}$ where $g^I$ are $m\times m$ matrices defined in (\ref{generator: Cartan subalgebra of su(m)}). In the $a_0=0$ gauge, $a_1$ and $a_2$ are conjugate variables since the Lagrangian density for internal gauge fields $a_\mu$ writes
\begin{eqnarray}\label{action:su(m)_1}
\notag\mathcal{L}_{SU(m)_1}[a]=\frac1{4\pi}\sum_{I=1}^{m-1}I(I+1)\Big(a^I_1\partial_t a^I_2-a^I_2\partial_t a^I_1\Big)
\end{eqnarray}
According to uncertainty principle, $a_1^I$ and $a_2^I$ cannot be determined simultaneously and we choose to fix the configuration of $a_2^I$. Aside from these $U(1)$ gauge symmetries, there are also discrete symmetries associated with essentially all permutations between partons (for details see Appendix \ref{app:g.s.deg}). Taking all these into account we can obtain the ground state degeneracy as the number of gauge-inequivalent configurations of $\{a_\mu^I\}$. As shown in Appendix \ref{app:g.s.deg}, the $m$-fold degenerate ground states correspond to the following gauge field configurations:
\begin{eqnarray}\label{gauge_configuration}
&\notag a_2^I=0,~(I=1,2,\cdots,m-2);\\
&a_2^{m-1}=\frac{2\pi}{L_2}\frac km,~~~k=1,\cdots,m.
\end{eqnarray}
Physically this means once we insert $2\pi k/m$ flux in the hole along $x_1$ direction of the torus for each parton, the original ground state is transformed into a different degenerate ground state. This is a ``small" gauge transformation for the partons since they transform as
\begin{eqnarray}
f_\alpha\rightarrow\exp\Big[\imth\sum_{i=1,2}x_2\sum_{I=1}^{m-1}a_2^Ig^I_{\alpha\beta}\Big]f_\beta
\end{eqnarray}

Now we add Higgs terms $M_{\alpha,\beta}({\bf r}|{\bf r}^\prime)$ which break the original $SU(m)$ gauge symmetry down to $Z_m$. Does the corresponding $Z_m$ state have the same ground state degeneracy as a $SU(m)$ state? The answer is positive. In the long-wavelength limit we introduce the Higgs fields $\phi_{\alpha\beta}$ which carry no electric $U(1)$ charge but carry the internal gauge charge. As an example, the $f_1^\dagger({\bf r})M_{1,m}({\bf r}|{\bf r}^\prime)f_m({\bf r}^\prime)$ terms in the lattice model will introduce Higgs field $\phi_{1,m}(x_1,x_2)$ in the long-wavelength limit. The Higgs field $\phi_{1,m}$ carries $a^I$ charge $+1$ for $I=1,\cdots,m-2$ and $a^{m-1}$ charge $+m$. Likewise, for example, $\phi_{2,m}$ carries $a^1$ charge $-1$, $a^I$ charge $+1$ for $I=2,\cdots,m-2$ and $a^{m-1}$ charge $+m$. In general for a Higgs field $\phi_{\alpha,\beta}=\phi^\ast_{\beta,\alpha}$,~$\alpha<\beta$ associated with mixing term $f^\dagger_\alpha f_\beta$ has $a^I_\mu$ charge $Q^I_{\alpha,\beta}$ where
\begin{eqnarray}\label{higgs:charge}
Q^I_{\alpha,\beta}=\left\{\begin{aligned}0,~~~&I\leq\alpha-2~~\text{or}~I\geq\beta\\
\alpha-1,~~~&I=\alpha-1\\
+1,~~~&\alpha\leq I\leq\beta-2\\
\beta,~~~&I=\beta-1\end{aligned}\right.
\end{eqnarray}
In the end one can see that the condensation of Higgs field can be viewed as adding a potential in the phase space of gauge field configurations to the $SU(m)_1$ Chern-Simons action (\ref{action:su(m)_1}). To be specific, once we integrate out partons with the presence of Higgs fields the effective Lagrangian density for internal gauge fields becomes
\begin{eqnarray}
&\notag\mathcal{L}_{eff}[a^I,\phi_{\alpha,\beta}]=\mathcal{L}_{SU(m)_1}[a]\\
&\notag+\mathcal{L}_{Higgs}\Big[(\partial_\mu-\imth\sum_IQ^I_{\alpha,\beta}a_\mu^I)\phi_{\alpha,\beta}\Big]
\end{eqnarray}
Note that the above action is invariant under the following ``large" gauge transformations
\begin{eqnarray}
&(a_1^I,a_2^I)\rightarrow(a_1^I+\frac{2\pi p_1}{L_1},a_2^I+\frac{2\pi p_2}{L_2}),\notag\\
&\phi_{\alpha,\beta}\rightarrow\phi_{\alpha,\beta}\exp\Big[\imth2\pi\sum_IQ^I_{\alpha,\beta}\big(\frac{p_1x_1}{L_1}+\frac{p_2x_2}{L_2}\big)\Big]\notag
\end{eqnarray}
where $p_{1,2}$ are integers so that $\phi_{\alpha,\beta}$ is a single-valued function on the torus. Besides there are other large gauge transformations as listed in (\ref{cs_other_large_g.t.}). And all the discrete gauge transformations associated with permutations between partons are also present, such as $P_{1,2}$
\begin{eqnarray}
&\notag f_1\longleftrightarrow f_2,~a^1_\mu\rightarrow-a_\mu^1,~\phi_{1,2}\longleftrightarrow\phi_{2,1}.
\end{eqnarray}
Upon integrating out the fluctuations of Higgs fields $\delta\phi_{\alpha,\beta}$ around their mean-field values $\bar\phi_{\alpha,\beta}$ in $\mathcal{L}_{eff}[a^I,\phi_{\alpha,\beta}]$ we have
\begin{eqnarray}\label{cshiggs}
\mathcal{L}_{Z_m}[a]=\mathcal{L}_{SU(m)_1}[a]-V[a^I_1,a^I_2]
\end{eqnarray}
The exact shape of potential $V[a_1^I,a_2^I]$ depends on \eg magnitudes of Higgs fields $\phi_{\alpha,\beta}$, but it has certain robust features determined by the gauge charges $Q^I_{\alpha,\beta}$ of Higgs fields $\phi_{\alpha,\beta}$ which condense\cite{Wen1990b}. More precisely, \emph{this potential is periodic in $a_\mu$ configuration space}, with periodicity $2\pi/L_1$ for $a_1^I,~I\leq m-2$ and $2\pi/L_2$ for $a_2^I,~I\leq m-2$. It also has periodicity $2\pi/(mL_1)$ for $a_1^{m-1}$ and $2\pi/(mL_2)$ for $a_2^{m-2}$. The minima of this potential sits exactly on the configurations shown in (\ref{gauge_configuration}) of the $m$-fold degenerate ground states of $SU(m)_1$ Chern-Simons theory. Besides these features associated with large gauge transformations, the potential $V[a^I_1,a^I_2]$ are not invariant under the discrete symmetries associated with parton permutations such as $P_{1,2}$. This is essentially because the introduced mixing terms (or Higgs condensation) breaks the $SU(m)$ gauge symmetry. The Lagrangian (\ref{cshiggs}) actually describes the motion of particles in a magnetic field\cite{Wen1998,Dunne1999} and a periodic potential: the $I$-th particle associated with $a_\mu^I$ experiences a magnetic field of $I(I+1)$ flux quanta piercing through the torus. Due to the periodicity of potential $V[a^I_1,a^I_2]$, the $m$-fold ground state degeneracy (as calculated in Appendix \ref{app:g.s.deg}) is still present when the gauge symmetry is broken from $SU(m)$ down to $Z_m$ by introducing mixing terms between different partons (or Higgs fields).

This can be understood physically: by threading a $2\pi k/m$ flux of gauge field $a^{m-1}_\mu$ in the hole along $x_1$ direction on the torus, one creates a vortex (or $2\pi$ phase winding) in the $\phi_{I,m},~I=1,\cdots,m-1$ condensates in the non-contractible loop along the $x_1$ direction. This operation exactly corresponds to the tunneling process $\mathcal{T}_2^k$ mentioned in section \ref{PARTON} and will cost zero energy in the thermodynamic limit. Therefore the presence of Higgs fields will not lift the $m$-fold ground state degeneracy in the thermodynamic limit.

We've shown that the ground state degeneracy of a $Z_m$ FCI state is $m$ on a torus. The above analysis for the torus case can be easily generalized to study the ground state degeneracy on a genus-$g$ Riemann surface. There are $g$ pairs of non-contractible loops $\{A_a,B_a|a=1,\cdots,g\}$ on a genus-$g$ Riemann surface where each pair is just like the two non-contractible loops on a torus. Thus one can straightforwardly show that the ground state degeneracy of a $Z_m$ FCI state is $m^g$ on a genus-$g$ Riemann surface.

\section{Parton construction of time-reversal-invariant FTI states}\label{FTI}

\subsection{$SU(m)^\uparrow\times SU(m)^\downarrow$ parton construction of TRI FTI states}

We have focused on spin-polarized FCI states. Taking into account spin degrees of freedom, nearly flat bands with time-reversal-invariant (TRI) $\mathbb{Z}_2$ index\cite{Sun2011} can exist. When a pair of bands carrying $\mathbb{Z}_2$ index are partially filled, can a fractionalized topological phase  preserving both time reversal and lattice symmetries be realized? In principle the answer is yes. As a direct generalization of spin-polarized $SU(m)$ and $Z_m$ FCI states, when the pair of nearly flat $\mathbb{Z}_2$ bands are filled partially with $\nu=2/m$, by $SU(m)^\uparrow\times SU(m)^\downarrow$ parton approach we can construct a TRI fractionalized phase which we term as $SU(m)^\uparrow\times SU(m)^\downarrow$ and $Z_m^\uparrow\times Z_m^\downarrow$ FTIs. In a simple way, a $SU(m)^\uparrow\times SU(m)^\downarrow$ ($Z_m^\uparrow\times Z_m^\downarrow$) FTI wavefunction with $\nu=2/m$ is a direct product of a spin-polarized $SU(m)$ ($Z_m$) FCI state with $\sigma_{xy}=e^2/(mh)$ for spin $\uparrow$ and its time-reversal counterpart: a spin-polarized $SU(m)$ ($Z_m$) FCI state with $\sigma_{xy}=-e^2/(mh)$ for spin $\downarrow$. Consequently in these FTI wavefunctions, entanglement between electrons with opposite spins is absent. We term these wavefunctions as spin-conserved FTI wavefunctions, and will construct FTI wavefunctions involving entanglement with opposite spins in the next subsection. But as a fully gapped topological phase, this phase is stable at least when the mixing between spin $\uparrow$ and spin $\downarrow$ is weak in the electronic Hamiltonian. As in the case of spin-polarized $SU(m)$ and $Z_m$ FCI states, we still use the technique of enlarging the unit cell by $m$ times, to guarantee that the ground state with the correct filling fraction is an insulator with a band gap.

The mean-field ansatz of a generic spin-conserved FTI wavefunction is written as
\begin{eqnarray}\label{mfH:fti:1/m}
&\notag H^{MF}_0=\sum_{{\bf r},{\bf r}^\prime}\sum_{\sigma,\sigma^\prime}f^\dagger_{\alpha,\sigma}({\bf r})\tilde M_{\alpha,\beta}({\bf r},\sigma|{\bf r}^\prime,\sigma^\prime)f_{\beta,\sigma^\prime}({\bf r}^\prime),\\
&\notag\tilde M_{\alpha,\beta}({\bf r},\sigma|{\bf r}^\prime,\sigma^\prime)=\\
&\delta_{\sigma,\sigma^\prime}\Big[M_{\alpha,\beta}({\bf r}|{\bf r}^\prime)\delta_{\sigma,\uparrow}+M^\ast_{\alpha,\beta}({\bf r}|{\bf r}^\prime)\delta_{\sigma,\downarrow}\Big]
\end{eqnarray}
where $\sigma,\sigma^\prime=\uparrow,\downarrow$ are the spin indices. $M_{\alpha,\beta}({\bf r}|{\bf r}^\prime)$ can be any mean-field ansatz of a $SU(m)$ (or $Z_m$) FCI state constructed in section \ref{PARTON} and demonstrated in section \ref{EXAMPLE}. And the corresponding FTI state is a $SU(m)^\uparrow\times SU(m)^\downarrow$ ($Z_m^\uparrow\times Z_m^\downarrow$) FTI. Apparently (\ref{mfH:fti:1/m}) is invariant under time-reversal transformation while preserving all the lattice symmetries.

Again the $N$-electron wavefunction (with $N/2$ electrons for each spin here in the $S^z$-conserved case) is obtained by projection on mean-field state $|MF\rangle=|MF_\uparrow\rangle\otimes|MF_\downarrow\rangle$
\begin{eqnarray}
\Phi_e(\{{\bf r}_i^\uparrow\};\{{\bf r}_j^\downarrow\})=\langle0|\prod_{i=1}^{N/2}c_\uparrow({\bf r}_i^\uparrow)\prod_{j=1}^{N/2}c_\downarrow({\bf r}_j^\downarrow)|MF\rangle,\label{eq:fti_electron}
\end{eqnarray}
which is simply a product of two Slater determinants (\ref{wavefunction:zm}) for spin up and down partons. Following our discussion in section \ref{EFT}, the ground state degeneracy of this $SU(m)^\uparrow\times SU(m)^\downarrow$ ($Z_m^\uparrow\times Z_m^\downarrow$) FTI is $m^2$ on a torus and there are quasiparticle excitations (with both spin) with electric charge $\pm e/m$.

 The $SU(m)^\uparrow\times SU(m)^\downarrow$ (or $Z_m^\uparrow\times Z_m^\downarrow$) FTI wavefunctions constructed above explicitly conserve the both the $\uparrow$ and $\downarrow$ electrons. However, in a spin-orbit coupled system where the $S_z$ conservation is not a symmetry, the true ground state wavefunction must involve mixings between the two spin species. Of course it is possible that this true ground state is in the same universality class as those spin-conserved FTI states, because they are gapped stable topological phases. Nevertheless it is still interesting to explicitly write down a FTI state without spin-conservation.

There is a natural question that needs to be answered in the current formalism: when spin mixing terms $f^\dagger_{\alpha,\uparrow}f_{\beta,\downarrow}$ are present in the parton mean-field Hamiltonian Eq.\ref{mfH:fti:1/m} while preserving the time-reversal symmetry, is the corresponding electronic state Eq.\ref{eq:fti_electron} a TRI FTI wavefunction without spin conservation? The answer is negative. Below we study the properties of this state in details.

The mean-field Hamiltonian including spin mixings between partons is
\begin{eqnarray}\label{eq:zm}
&\notag H^{MF}=H_0^{MF}+\sum_{\alpha\beta}\big[f^\dagger_{\alpha,\uparrow}({\bf r})\tilde M_{\alpha,\beta}({\bf r},\uparrow|{\bf r}^\prime,\downarrow)f_{\beta,\downarrow}({\bf r}^\prime)\\
&+~h.c.\big]
\end{eqnarray}
Note that upon mixing partons with different spins, \eg for a $Z_m^\uparrow\times Z_m^\downarrow$ FTI the internal gauge symmetry is further broken from $Z_m^\uparrow\times Z_m^\downarrow$ (one $Z_m^\uparrow$ for spin-$\uparrow$ parton and another independent $Z_m^\downarrow$ for spin-$\downarrow$ parton)
\begin{eqnarray}\label{igg:tfi:zmxzm}
f_{\alpha,\sigma}({\bf r})\rightarrow e^{\imth\frac{2\pi a_\sigma}{m}}f_{\alpha,\sigma}({\bf r}),~~~a_{\uparrow},a_{\downarrow}=1,2,\cdots,m.
\end{eqnarray}
to a single $Z_m$ symmetry (for partons with both spin $\uparrow$ and $\downarrow$)\footnote{For a $SU(m)^\uparrow\times SU(m)^\downarrow$ FTI, the internal gauge symmetry is further broken down to an overall $SU(m)$ by mixing of partons with different spins but the same flavor. The unbroken $SU(m)$ gauge field does not have a Chern-Simons term. Consequently, this state suffers from the well-known confinement problem of the dynamical $SU(m)$ gauge field and does not have a stable mean-field description. We therefore do not discuss this state in this paper.}:
\begin{eqnarray}\label{igg:tfi:zm}
f_{\alpha,\sigma}({\bf r})\rightarrow e^{\imth\frac{2\pi a}{m}}f_{\alpha,\sigma}({\bf r}),~~~a=1,2,\cdots,m.
\end{eqnarray}

In this paper, we define the FTI phase as the phase of matter which is in the same universality class as the direct product of two FCI states of opposite spins while preserving the time-reversal symmetry. Following this definition, the $Z_m$ state defined in Eq.\ref{eq:zm} with spin mixings is \emph{not} a FTI state.

One clear difference between the $Z_m$ state Eq.\ref{eq:zm} and the $Z_m^\uparrow\times Z_m^\downarrow$ FTI state is their quasiparticle statistics. In a $Z_m^\uparrow\times Z_m^\downarrow$ FTI state, the $Z_m^\uparrow$ fluxes and the $Z_m^\downarrow$ fluxes bind with parton charges due to the Chern numbers of the parton bands, and are anyons with statistical angles $\theta=\pm\frac{\pi}{m}$. But in a $Z_m$ state Eq.\ref{eq:zm}, the $Z_m$ fluxes do not bind with parton charges, and are bosons. This indicates that the $Z_m$ state Eq.\ref{eq:zm}, which is described by a regular $Z_m$ lattice gauge dynamics with bosonic flux excitations, must separate from the $Z_m^\uparrow\times Z_m^\downarrow$ FTI state by a phase transition.

Another observation that we have is that, in the $SU(m)^\uparrow\times SU(m)^\downarrow$ parton construction, the $Z_m$ state Eq.\ref{eq:zm} cannot preserves both time-reversal symmetry and lattice translation symmetry simultaneously. Essentially the technique of enlarging unit cell $m$ times by inserting fluxes fails to generate a translation symmetric wavefunction when partons with opposite spins are mixed. In a TRI $Z_m^\uparrow\times Z_m^\downarrow$ FTI state, when $2\pi/m$ fluxes are inserted in each unit cell for spin $\uparrow$ partons, an opposite $-2\pi/m$ fluxes are inserted in each unit cell for spin $\downarrow$ parton to preserve the time-reversal symmetry. After the partons with opposite spins are mixed, the gauge flux pattern of the partons no longer enjoys well-defined $\pm 2\pi/m$ value per plaquette. As a result physical translation symmetry is broken.
This simple argument dictates that, \emph{using $SU(m)^\uparrow\times SU(m)^\downarrow$ parton construction, when partons with opposite spins are mixed, either time reversal symmetry or lattice translation symmetry must be broken to form a gapped state with filling fraction $2/m$ ($\nu=1/m$ for each spin on average)}. In Appendix \ref{app:psg:fti:checkerboard} we prove this statement by a careful PSG analysis.

The analysis in this subsection seems to suggest that it is difficult to explicitly construct a FTI wavefunctions breaking the $S_z$ conservation. In fact, this difficulty is due to the formalism of $SU(m)^\uparrow\times SU(m)^\downarrow$ parton construction. In the next subsection, we propose another parton construction formalism which allows us to explicitly write down FTI wavefunctions with spin mixings.

\subsection{Parton construction of generic TRI FTI states in the absence of spin conservation}

In the following we demonstrate that in a new parton construction formalism, one can write down the electron wavefunctions for fully symmetric TRI FTI states breaking the $S_z$ conservation, and with mean-field terms mixing partons with different spins. We introduce the following parton construction ($m$ being an odd integer and $\theta$ is an arbitrary real number)
\begin{eqnarray}
&c_\uparrow({\bf r})=\cos\theta\prod_{\alpha=1}^mf_{\alpha,\uparrow}({\bf r})+\sin\theta\prod_{\beta=1}^mg_{\beta,\uparrow}({\bf r}),\notag\\
&c_\downarrow({\bf r})=-\sin\theta\prod_{\alpha=1}^mf_{\alpha,\downarrow}({\bf r})+\cos\theta\prod_{\beta=1}^mg_{\beta,\downarrow}({\bf r}).\notag\\
\label{electron:tri}
\end{eqnarray}
where $f_{\alpha,\sigma}$ and $g_{\beta,\sigma}$ are all fermionic partons each of which carries electric charge $e/m$. It's straightforward to see that the electron constructed in this way is indeed a fermion with electric charge $e$. The $N$-electron wavefunction at filling fraction $\nu=2/m$ (with $N_\uparrow$ spin-$\uparrow$ electrons and $N_\downarrow=N-N_\uparrow$ spin-$\downarrow$ electrons) is obtained by projection
\begin{eqnarray}\label{wavefunction:tri}
\Phi_e(\{{\bf r}_{i}^\uparrow\};\{{\bf r}_{j}^{\downarrow}\})=\langle0|\prod_{i=1}^{N_\uparrow}c_\sigma({\bf r}_{i}^{\uparrow})\prod_{j=1}^{N-N_\uparrow}c_\sigma({\bf r}_{j}^{\downarrow})|MF\rangle
\end{eqnarray}
where $|MF\rangle$ is the parton mean-field ground state as will be described later. $N_\uparrow$ can be any integer between $0$ and total electron number $N$.

The mean-field ansatz can be written as
\begin{eqnarray}
&\notag H^{MF}=\sum_{\alpha,\alpha^\prime=1}^m\sum_{\sigma,\sigma^\prime=\uparrow,\downarrow}\sum_{{\bf r},{\bf r}^\prime}\\
\notag&\Big(f^\dagger_{\alpha,\sigma}({\bf r})M_{\alpha,\alpha^\prime}({\bf r},\sigma|{\bf r}^\prime,\sigma^\prime)f_{\alpha^\prime,\sigma^\prime}({\bf r}^\prime)\\
&+g^\dagger_{\alpha,\sigma}({\bf r})M^\prime_{\alpha,\alpha^\prime}({\bf r},\sigma|{\bf r}^\prime,\sigma^\prime)g_{\alpha^\prime,\sigma^\prime}({\bf r}^\prime)\Big)\label{mfH:tri_mixed}
\end{eqnarray}
where Hamiltonian $M^\prime_{\alpha,\alpha^\prime}({\bf r},\sigma|{\bf r}^\prime,\sigma^\prime)$ is the time reversal conjugate of $M_{\alpha,\alpha^\prime}({\bf r},\sigma|{\bf r}^\prime,\sigma^\prime)$:
\begin{eqnarray}
M^\prime_{\alpha,\alpha^\prime}({\bf r},\sigma|{\bf r}^\prime,\sigma^\prime)=\sigma\sigma^\prime M^\ast_{\alpha,\alpha^\prime}({\bf r},-\sigma|{\bf r}^\prime,-\sigma^\prime)
\end{eqnarray}
We use spin index $\sigma=\pm1$ to denote spin $\uparrow,\downarrow$. In the simplest case when $M_{\alpha,\alpha^\prime}({\bf r},\sigma|{\bf r}^\prime,\sigma^\prime)=\delta_{\alpha,\alpha^\prime}M({\bf r},\sigma|{\bf r}^\prime,\sigma^\prime)$, the mean-field ansatz has $SU(m)$ gauge symmetry or in other words $IGG=SU(m)$. As discussed earlier in section \ref{PARTON}, in a generic case parton mixing terms $f^\dagger_{\alpha,\sigma}f_{\beta,\sigma^\prime},\alpha\neq\beta$ could exist and the IGG of the parton mean-field Hamiltonian $M_{\alpha,\alpha^\prime}({\bf r},\sigma|{\bf r}^\prime,\sigma^\prime)$ could be $Z_m$, which is the center of group $SU(m)$. Here $M_{\alpha,\alpha^\prime}({\bf r},\sigma|{\bf r}^\prime,\sigma^\prime)$ can be any mean-field ansatz as a solution of the PSG constraints on the lattice, so that the $N$-electron wavefunction obtained by projection (\ref{wavefunction:tri}) preserves all the lattice symmetries. More precisely, mean-field Hamiltonian $M_{\alpha,\alpha^\prime}({\bf r},\sigma|{\bf r}^\prime,\sigma^\prime)$ should be invariant under a symmetry operation $U$ followed by a $SU(m)$ gauge rotation $G_U({\bf r},\sigma)$ on $f$-partons.

For instance, the lattice symmetry groups are shown in Appendix \ref{app:sym:honeycomb} for honeycomb lattice and in Appendix \ref{app:sym:checkerboard} for checkerboard lattice for spin $\uparrow/\downarrow$. The constraints for $G_U({\bf r},\sigma)$ will still be those in Appendix \ref{app:psg:honeycomb} on honeycomb lattice and Appendix \ref{app:psg:checkerboard} on checkerboard lattice in the case when $IGG=Z_m$. Namely $G_U({\bf r},\uparrow)$ and $G_U({\bf r},\downarrow)$ can be any two solutions of PSG constraints: (\ref{psg:honeycomb}) on honeycomb lattice and (\ref{psg:checkerboard}) on checkerboard lattice when $IGG=Z_m$. The symmetry-allowed mean-field Hamiltonian $M_{\alpha,\alpha^\prime}({\bf r},\sigma|{\bf r}^\prime,\sigma^\prime)$ can be obtained in the same way shown in Appendix \ref{app:psg:honeycomb} and Appendix \ref{app:psg:checkerboard}. It's easy to check that spin-mixing terms in mean-field Hamiltonian $M_{\alpha,\alpha^\prime}({\bf r},\sigma|{\bf r}^\prime,\sigma^\prime)$ is in general allowed by these PSG constraints.

Meanwhile time reversal symmetry is also preserved since the anti-unitary time reversal operation $\bst$ is realized by complex conjugation $\mathcal{C}$ combined with the following operation
\begin{eqnarray}
f_{\alpha,\sigma}({\bf r})\leftrightarrow\sigma g_{\alpha,-\sigma}({\bf r})
\end{eqnarray}
where spin index $\sigma=\pm1$ denotes spin $\uparrow/\downarrow$. One can easily check that under time reversal $\bst$ the spin-$1/2$ electron operators indeed transform as $c_\sigma({\bf r})\rightarrow\sigma c_{-\sigma}({\bf r})$. Apparently the above time reversal operation $\bst$ leaves the mean-field ansatz (\ref{mfH:tri_mixed}) invariant. At filling $\nu=2/m$ ($1/m$ filling for each spin \emph{on average}) we still use the technique of inserting flux to enlarge the unit cell by $m$ times. Note that when $2\pi/m$ flux is inserted into each unit cell for $f$-partons, an opposite $-2\pi/m$ flux must be inserted in each unit cell for $g$-partons to keep the time reversal symmetry. Then we fill the lowest $m$ bands for both $f$-partons and $g$-partons. Each filled band contains $N/2$ $f$-partons (or $g$-partons) which correspond to a filling fraction of $1/m$. It's straightforward to demonstrate that the electron filling fraction of state (\ref{wavefunction:tri}) is indeed $\nu=2/m$. There can be symmetry-allowed mixing terms between partons with different spins in ansatz $M_{\alpha,\alpha^\prime}({\bf r},\sigma|{\bf r}^\prime,\sigma^\prime)$. The mean-field state in (\ref{wavefunction:tri}) is a direct product of $f$-parton state and $g$-parton state: $|MF\rangle=|MF^f\rangle\otimes|MF^g\rangle$.

There is a real parameter $\theta\in[0,\pi/4]$ which can be continuously tuned in our parton construction (\ref{electron:tri}). This parameter controls the many-body entanglement between spin-$\uparrow$ and spin-$\downarrow$ electrons in wavefunction (\ref{wavefunction:tri}). It should be considered as a variational parameter in variational Monte Carlo studies of projected wavefunctions. When $\theta=0$ clearly there must be equal number of spin-$\uparrow$ and spin-$\downarrow$ electrons: $N_\uparrow=N_\downarrow=N/2$ since other components of the many-body wavefunction with $N_\uparrow\neq N_\downarrow$ all vanish in (\ref{wavefunction:tri}). In this case the electron wavefunction (\ref{wavefunction:tri}) is nothing but a direct product of spin-$\uparrow$ wavefunction $\Phi_\uparrow({\bf r}_i^\uparrow)=\langle0|\prod_{i=1}^{N/2}\prod_{\alpha=1}^mf_{\alpha,\uparrow}({\bf r}_i^\uparrow)|MF^f\rangle$ and spin-$\downarrow$ wavefunction $\Phi_\downarrow({\bf r}_j^\downarrow)=\langle0|\prod_{j=1}^{N/2}\prod_{\alpha=1}^mg_{\alpha,\downarrow}({\bf r}_j^\downarrow)|MF^g\rangle$. This corresponds to the spin-conserved limit when there is no entanglement between electrons with different spins.

When $\theta$ is nonzero, many-body entanglement between electrons with different spins is encoded in electron wavefunction (\ref{wavefunction:tri}) as long as the spin mixing terms are present in the mean-field Hamiltonian (\ref{mfH:tri_mixed}). And in general the component of the many-body wavefunction with an arbitrary number of spin-$\uparrow$ electrons $\forall~0\leq N_\uparrow\leq N$ should be nonzero. In a generic case with $\theta\neq0$ the many-body wavefunction (\ref{wavefunction:tri}) is complicated and cannot be written as a Slater determinant. Now one can see the parton construction (\ref{electron:tri}) allows us to write down generic electron wavefunctions for TRI FTI states in the absence of spin conservation. The spin-conserved TRI FTI wavefunction ($\theta=0$) can be deformed into a generic TRI FTI state in the absence of spin conservation ($\theta\neq0$) by continuously tuning parameter $\theta$, while keeping the mean-field Hamiltonian (\ref{mfH:tri_mixed}) unchanged. In the process of tuning $\theta$ continuously, we expect the low-energy effective theory and quasiparticles of such a TRI FTI state to remain the same.

In the end we comment on the low-energy effective theory of such a TRI FTI state. In the simplest case when $M_{\alpha,\alpha^\prime}({\bf r},\sigma|{\bf r}^\prime,\sigma^\prime)=\delta_{\alpha,\alpha^\prime}M({\bf r},\sigma|{\bf r}^\prime,\sigma^\prime)$, the mean-field ansatz (\ref{mfH:tri_mixed}) has a $SU(m)\times SU(m)$ gauge symmetry or in other words $IGG=SU(m)^f\times SU(m)^g$. Let's assume the filled lowest band of $M({\bf r},\sigma|{\bf r}^\prime,\sigma^\prime)$ for $\{f_{\alpha,\uparrow/\downarrow}|\alpha=1,\cdots,m\}$ partons has a Chern number $k$. Then due to time reversal symmetry the filled lowest band for $\{g_{\alpha,\uparrow/\downarrow}|\alpha=1,\cdots,m\}$ partons will have a Chern number $-k$. And its low-energy effective theory is a $SU(m)_k\times SU(m)_{-k}$ Chern-Simons theory:
\begin{eqnarray}
&\notag\mathcal{L}_{eff}=\frac k{4\pi}\epsilon_{\mu\nu\lambda}\text{Tr}\big(a_\mu\partial_\nu a_\lambda+\frac\imth3a_\mu a_\nu a_\lambda\big)\\
&-\frac k{4\pi}\epsilon_{\mu\nu\lambda}\text{Tr}\big(b_\mu\partial_\nu b_\lambda+\frac\imth3b_\mu b_\nu b_\lambda\big)\label{eff:tri su(m)}
\end{eqnarray}
where $a_\mu$ is the $SU(m)$ gauge field coupled to $f$-partons and $b_\mu$ is the $SU(m)$ gauge field coupled to $g$-partons. Such a $SU(m)^f\times SU(m)^g$ TRI FTI state will host non-Abelian quasiparticles if $k>1$.

When $k=1$ this is an Abelian TRI FTI state with ground state degeneracy $m^2$ on a torus and anyonic quasiparticles. When parton mixing terms $f^\dagger_{\alpha,\sigma}f_{\beta,\sigma^\prime},~\alpha\neq\beta$ are present, again the IGG is reduced from $SU(m)^f\times SU(m)^g$ to ${Z_m}^f\times{Z_m}^g$ and the low-energy effective theory is described by Chern-Simons-Higgs theory, \ie effective action (\ref{eff:tri su(m)}) with a periodic potential due to Bose condensation of Higgs fields. Such an Abelian ${Z_m}^f\times{Z_m}^g$ TRI FTI has the same topological properties as an Abelian $SU(m)^f\times SU(m)^g$ FTI, such as ground state degeneracy and quasiparticle charge/statistics. In the parton construction (\ref{electron:tri}), both $SU(m)^f\times SU(m)^g$ and ${Z_m}^f\times{Z_m}^g$ TRI FTI states are possible candidates for a symmetric TRI FTI state in the absence of spin conservation: which state is realized depends on the IGG of mean-field amplitudes $M_{\alpha,\alpha^\prime}({\bf r},\sigma|{\bf r}^\prime,\sigma^\prime)$ and should be determined by energetics of wavefunctions (\ref{wavefunction:tri}) in variational studies.

In the end we comment on the robustness of such a symmetry-protected TRI FTI state against perturbations. As discussed in the previous subsection, in the presence of time reversal symmetry and lattice translation symmetry, no terms that mix $f_{\uparrow/\downarrow}$ partons and $g_{\uparrow/\downarrow}$ partons can be added to mean-field ansatz (\ref{mfH:tri_mixed}). Since both $f$-partons and $g$-partons form a band insulator, such a TRI FTI is stable in the absence of mean-field terms mixing $f$- and $g$-partons. On the other hand, once time reversal symmetry or translation symmetry is broken one can always add a mixing term in the mean-field ansatz (\ref{mfH:tri_mixed}), which will drive the system from a FTI state into a different phase. Therefore the stability of this TRI FTI state is protected by time reversal symmetry.

\section{Conclusion}

To summarize, we show that a large class of Abelian (and non-Abelian) fractionalized topological phases can be constructed on a lattice using parton construction. These states preserve all the lattice symmetry and are featured by \eg fractionalized excitations and topological ground state degeneracy. In the spin-polarized case, we construct  fractional Chern insulator phases belong to distinct universality classes even at the same filling $\nu=1/m$. Their differences are characterized by the projective symmetry group in the bulk and are protected by the lattice symmetry. The low energy gauge groups of these states are found to be either $SU(m)$ or $Z_m$. Their low-energy physics is described by $SU(m)_1$ Chern-Simons theory and Chern-Simons-Higgs theory respectively, and they all have $m$-fold degenerate ground states on a torus. We explicitly construct the ground state wavefunctions and bulk quasiparticles on the lattice. We demonstrate our construction by several explicit examples, including non-Abelian FCIs which may be realized in a nearly flat band with Chern number $C>1$. Furthermore, we show that when time reversal symmetry is present, classes of fractionalized topological insulator phases preserving both time reversal symmetry and lattice symmetries can be constructed. These TRI FTI states are characterized by $SU(m)\times SU(m)$ or $Z_m\times Z_m$ gauge groups. Their electron wavefunctions on the lattice, which are essentially products of spin-polarized FCI states for spin $\uparrow$ and its time reversal conjugate, are provided. These are stable topological phases even when $S_z$ conservation is not a symmetry in the electronic system. In order to explicitly construct TRI FTI wavefunctions with entanglement between opposite spins, we propose a new parton construction formalism. It allows one to write down generic electron wavefunctions of TRI FTI states, which preserve both time reversal and lattice symmetries in the absence of spin conservation. Our work provides important insight for future numeric study using variational Monte Carlo method.

We thank Ashvin Vishwanath and Xiao-Gang Wen for helpful comments. YML is indebted to Yi Zhang for sharing some unpublished numerical results. YML is supported by DOE grant
DE-FG02-99ER45747. YR is supported by start-up fund at Boston College.

After this manuscript was submitted, we note that another paper\cite{McGreevy2011} also discusses the technique of inserting $2\pi/m$ fluxes in unit cells in order to construct translational invariant FCI wavefunctions by parton approach.

\appendix

\section{Symmetry group of the honeycomb lattice model}\label{app:sym:honeycomb}

The symmetry group of Haldane's model is generated by the following symmetry operations as shown in FIG. \ref{fig:lattice}(a):
(1) translations $T_{1,2}$ by Bravais lattice vector $\vec a_{1,2}$;
(2) $\pi/3$ rotation $\cs$ along $\hat z$-axis around the honeycomb plaquette center;
(3) mirror reflection w.r.t. $\hat x$-$o$-$\hat z$ plane combined with time-reversal operation, labeled as $\bss$.

Note that $\bss$ is an anti-unitary symmetry since it includes time-reversal operation. It acts on the Hamiltonian through a combination of a unitary symmetry operation and complex conjugation $\mathcal{C}$. Besides, the spatial $C_6$ rotation should be accompanied with a corresponding spin rotation along $S^z$-axis in a generic electron Hamiltonian with spin-orbit coupling terms.

We label a lattice site by coordinates $(x,y,s)$ where $\vec r=x\vec a_1+y\vec a_2+\vec r_s$ is its position vector. $\vec a_1=a(\sqrt{3},0)$ and $\vec a_2=a(\sqrt{3},3)/2$ are two Bravais lattice vectors. $s=0,1$ is the sublattice index with $\vec r_0=-a(\sqrt{3},1)/2$ and $\vec r_1=a(-\sqrt{3},1)/2$. Under the symmetry operations the $(x,y,s)$ coordinates transform as
\begin{eqnarray}
\label{symmetry:honeycomb} T_1:~~~&(x,y,s)\rightarrow(x+1,y,s)\\
\notag T_2:~~~&(x,y,s)\rightarrow(x,y+1,s)\\
\notag \bss:~~~&(x,y,s)\rightarrow(x+y,-y,1-s)\\
\notag \cs:~~~&(x,y,s)\rightarrow(1-s-y,x+y+s-1,1-s)
\end{eqnarray}

The multiplication rules of the above symmetry group are completely determined by the following algebraic relations:
\begin{eqnarray}
&\label{algebra:honeycomb_fqh} \cs^6=\bss^2=\bse,\\
&\notag T_1^{-1}T_2T_1T_2^{-1}=\bse,\\
&\notag T_2^{-1}\cs T_1\cs^{-1}=\bse,\\
&\notag T_1^{-1}\cs T_1T_2^{-1}\cs^{-1}=\bse,\\
&\notag T_1^{-1}\bss T_1\bss^{-1}=\bse,\\
&\notag T_2^{-1}\bss T_1T_2^{-1}\bss^{-1}=\bse,\\
&\notag \bss\cs\bss\cs=\bse.
\end{eqnarray}
where $\bse$ represents the identity element of the symmetry group.

\section{Mean-field ansatz of spin-polarized $Z_m$ FCI states on honeycomb lattice in $SU(m)$ parton construction}\label{app:psg:honeycomb}

\subsection{Projective symmetry group (PSG) analysis of different $Z_m$ mean-field states}

Since the IGG of a spin-polarized $Z_m$ state is
\begin{eqnarray}
G_\bse\in Z_m=\{\eta\cdot I_{m\times m}|\eta=e^{\imth 2\pi\frac km},~~k=1,\cdots,m\}
\end{eqnarray}
the algebraic relations (\ref{algebra:honeycomb_fqh}) give us algebraic conditions on the gauge transformations $G_U(x,y,s)$:
\begin{eqnarray}
&G_{\bss}\big(\bss({\bf r})\big)G_{\bss}^\ast({\bf r})=\eta_\bss I_{m\times m},\label{sig^2}\\
&G_{C_6}\big(C_6^{-1}({\bf r})\big)G_{C_6}\big(C_6^{-2}({\bf r})\big)G_{C_6}\big(C_6^{3}({\bf r})\big)\\
&\cdot
G_{C_6}\big(C_6^{2}({\bf r})\big)G_{C_6}\big(C_6({\bf r})\big)G_{C_6}({\bf r})=\eta_{C_6}I_{m\times m},\label{c6^6}\\
&G_{T_1}^{-1}\big(T_2^{-1}T_1({\bf r})\big)G_{T_2}^{-1}\big(T_1({\bf r})\big)\notag\\
&\cdot G_{T_1}\big(T_1({\bf r})\big)G_{T_2}({\bf r})=\eta_{12}I_{m\times m},\label{T1-T2}\\
&G_{T_2}^{-1}\big(T_2({\bf r})\big)G_{C_6}\big(T_2({\bf r})\big)\notag\\
&\cdot G_{T_1}\big(T_1C_6^{-1}({\bf r})\big)G_{C_6}^{-1}({\bf r})=\eta_{C_61}I_{m\times m},\label{T2-c6}\\
&G_{T_1}^{-1}\big(T_1({\bf r})\big)G_{C_6}\big(T_1({\bf r})\big)G_{T_1}\big(C_6^{-1}T_1({\bf r})\big)\notag\\
&\cdot G_{T_2}^{-1}\big(C_6^{-1}({\bf r})\big)G_{C_6}^{-1}({\bf r})=\eta_{C_62}I_{m\times m},\label{T1-c6}\\
&G_{T_1}^{-1}\big(T_1({\bf r})\big)G_{\bss}\big(T_1({\bf r})\big)\notag\\
&\cdot G_{T_1}^\ast\big(T_1\bss^{-1}({\bf r})\big)G_{\bss}^{-1}({\bf r})=\eta_{\bss1}I_{m\times m},\label{T1-sig}\\
&G_{T_2}^{-1}\big(T_2({\bf r})\big)G_{\bss}\big(T_2({\bf r})\big)G_{T_1}^\ast\big(\bss T_2({\bf r})\big)\notag\\
&\cdot \big[G_{T_2}^{-1}\big(\bss({\bf r})\big)\big]^\ast G_{\bss}^{-1}({\bf r})=\eta_{\bss2}I_{m\times m},\label{T2-sig}\\
&G_{\bss}({\bf r})G_{C_6}^\ast\big(\bss({\bf r})\big)\notag\\
&\cdot G_{\bss}^\ast\big(\bss C_6({\bf r})\big)G_{C_6}\big(C_6({\bf r})\big)=\eta_{\bss C_6}I_{m\times m},\label{c6-sig}
\end{eqnarray}
where all the $\eta$'s are $Z_m$ quantum numbers $\{e^{\imth 2\pi\frac km}|k=1,\cdots,m\}$. Note that $\bss$ is a anti-unitary symmetry so it's accompanied with complex conjugate $\mathcal{C}$. We can always choose a proper gauge so that $G_{T_1}(x,y,s)=G_{T_2}(0,y,s)=I_{m\times m}$. $\eta_{\cs1}=\eta_{\cs2}=1$ can also be fixed by certain gauge choice. After solving the above algebraic conditions we have $\eta_{\bss1}=\eta_{\bss2}=\eta_{12}^{-1}$ and $\eta_{\bss}=1$. There are only two independent $Z_m$ quantum numbers left, \ie $\eta_{12}$ and $\eta_{\bss\cs}$. In the end we have
\begin{eqnarray}
&\label{psg:honeycomb}G_{T_1}(x,y,s)=I_{m\times m},~~~G_{T_2}(x,y,s)=\eta_{12}^xI_{m\times m},\\
&\notag G_\bss(x,y,s)=\eta_{12}^{-x-y(y+1)/2}g_\bss(s),\\
&\notag G_\cs(x,y,s)=\eta_{12}^{xy+x(x-1)/2}g_\cs(s),\\
&\notag \big[g_\cs(0)g_\cs(1)\big]^3=\eta_{12}\eta_\cs,~~~g_\bss(1)g^\ast_\bss(0)=1,\\
&\notag g_\cs(s)g_\bss(1-s)g^\ast_\cs(s)g^\ast_\bss(1-s)=\eta_{\bss\cs}.
\end{eqnarray}
where $g_\bss(s)$ and $g_\cs(s)$ are all $SU(m)$ matrices.

Different $Z_m$ mean-field states are given by gauge-inequivalent solutions of equations (\ref{psg:honeycomb}).

\subsection{Symmetry conditions on $Z_m$ mean-field states}

To choose a representative of all mean-field amplitudes $M({\bf r}|{\bf r}^\prime)$, we choose ${\bf r}^\prime=(0,0,0)$ and label independent mean-field bonds as
\begin{eqnarray}
[x,y,s]\equiv M(x,y,s|0,0,0)
\end{eqnarray}
All other symmetry-related mean-field amplitudes can be generated by $[x,y,s]$: \eg all 1st nearest neighbor (n.n.) mean-field amplitudes can be generated by $[0,0,1]$ through symmetry operations; all 2nd n.n. amplitudes can be generated by $[0,1,0]$.

Using symmetry operations (\ref{symmetry:honeycomb}) it's straightforward to verify the following relations
\begin{eqnarray}
&\notag T_2^{-1}\bss C_6:~~[-2x,x,0]\rightarrow[
-2x,x,0]\\
&T_2^{x-1}\bss C_6:~~[0,x,0]\rightarrow[0,x,0]^\dagger\notag\\
&T_1^{-1}\bss C_6^3:~~[x,-2x,0]\rightarrow[x,-2x,0]\notag\\
&T_1^{x-1}\bss C_6^3:~~[x,0,0]\rightarrow[x,0,0]^\dagger\notag\\
&\bss C_6^{-1}:~~[x,x,0]\rightarrow[x,x,0]\notag\\
&T_1^{x}T_2^{-x}\bss C_6^{-1}:~~[x,-x,0]\rightarrow[x,-x,0]^\dagger\notag
\end{eqnarray}
for mean-field amplitudes $[x,y,0]=M(x,y,0|0,0,0)$ between sites

within the same sublattice and
\begin{eqnarray}
&\notag T_1^xT_2^{-2x}\bss:~~[x,-2x,1]\rightarrow[
x,-2x,B]^\dagger\\
&\bss C_6^{-1}:~~[x+1,x,1]\rightarrow[
x+1,x,B]\notag\\
&T_1^{-2x-2}T_2^{x+1}\bss C_6^{-2}:~~[-2x-1,x,1]\rightarrow[-2x-1,x,1]^\dagger\notag\\
&T_1^{x-1}T_2^{y}C_6^3:~~[x,y,1]\rightarrow[x,y,1]^\dagger\notag
\end{eqnarray}
for mean-field amplitudes $[x,y,1]=M(x,y,1|0,0,0)$ between sites
from different sublattices. The symmetry condition (\ref{mfH:symmetry}) now becomes constraints on the mean-field amplitudes $[x,y,s]$.

In the following we list the symmetry conditions on mean-field amplitudes for 1st n.n. $u_\alpha\equiv[0,0,1]$
\begin{eqnarray}
&\notag g_\bss(1)u_\alpha^\ast g_\bss^\dagger(0)=u_\alpha^\dagger,\\
&\notag g_\cs(0)g_\cs(1)g_\cs(0)u_\alpha g_\cs^\dagger(1)g_\cs^\dagger(0)g_\cs^\dagger(1)=u_\alpha^\dagger.
\end{eqnarray}
and 2nd n.n. $u_\beta\equiv[0,1,0]$
\begin{eqnarray}
\notag g_\bss(0)g_\cs^\ast(1)u_\beta^\ast g_\cs^T(1)g_\bss^\dagger(0)=u_\beta^\dagger.
\end{eqnarray}

\section{Symmetry group of the checkerboard lattice model}\label{app:sym:checkerboard}

The symmetry group of checkerboard lattice model is generated by the following symmetry operations as shown in FIG. \ref{fig:lattice}(b):
(1) translations $T_{1,2}$ by Bravais lattice vector $\vec a^\prime_{1,2}$;
(2) mirror reflection w.r.t. $\hat x$-$o$-$\hat z$ plane combined with time-reversal operation, labeled as $P_x$.
(3) mirror reflection w.r.t. $\hat y$-$o$-$\hat z$ plane combined with time-reversal operation, labeled as $P_y$.
(4) mirror reflection along $(\hat x+\hat y)$-$o$-$\hat{z}$ plane combined with time-reversal operation, labeled as $\pxy$.

Again here $P_x,~P_y$ and $\pxy$ are all anti-unitary symmetries. They act on the Hamiltonian through a combination of unitary symmetry operations and complex conjugation $\mathcal{C}$.

Here we also label each lattice site by coordinates $(x,y,s)$ where $\vec r=x\vec a_1+y\vec a_2+\vec r_s$ corresponds to its position vector. $\vec a_1=a(1,0)$ and $\vec a_2=a(0,1)$ are two Bravais lattice vectors and $a$ is the lattice constant. Again $s=0,1$ is the sublattice index with $\vec r_0=-a(1/2,0)$ and $\vec r_1=-a(0,1/2)$. Under the symmetry operations the $(x,y,s)$ coordinates transform as
\begin{eqnarray}
\label{symmetry:checkerboard} T_1:~~~&(x,y,s)\rightarrow(x+1,y,s)\\
\notag T_2:~~~&(x,y,s)\rightarrow(x,y+1,s)\\
\notag P_x:~~~&(x,y,s)\rightarrow(x,s-y,s)\\
\notag P_y:~~~&(x,y,s)\rightarrow(1-s-x,y,s)\\
\notag \pxy:~~~&(x,y,s)\rightarrow(y,x,1-s)
\end{eqnarray}

The multiplication rules of the symmetry group are completely determined by the following algebraic relations:
\begin{eqnarray}
&\label{algebra:checkerboard_fqh} P_x^2=P_y^2=\pxy^2=\bse,\\
&\notag T_1^{-1}T_2T_1T_2^{-1}=\bse,\\
&\notag T_1^{-1}P_x T_1P_x^{-1}=\bse,\\
&\notag T_2^{-1}P_x T_2^{-1}P_x^{-1}=\bse,\\
&\notag T_1^{-1}P_y T_1^{-1}P_y^{-1}=\bse,\\
&\notag T_2^{-1}P_y T_2P_y^{-1}=\bse,\\
&\notag T_1^{-1}\pxy T_2\pxy^{-1}=\bse,\\
&\notag T_2^{-1}\pxy T_1\pxy^{-1}=\bse,\\
&\notag P_x P_y P_x P_y=\bse,\\
&\notag \pxy P_y\pxy P_x=\bse,\\
&\notag \pxy P_x\pxy P_y=\bse.
\end{eqnarray}
where $\bse$ represents the identity element of the symmetry group.

\section{Mean-field ansatz of spin-polarized $Z_m$ FCI states on checkerboard lattice in $SU(m)$ parton construction}\label{app:psg:checkerboard}

\subsection{Projective symmetry group (PSG) analysis of different $Z_m$ mean-field states}

From algebraic relations (\ref{algebra:checkerboard_fqh}) we have algebraic conditions on the gauge transformations $G_U(x,y,s)$:
\begin{eqnarray}
&G_{T_1}^{-1}\big(x+1,y-1,s\big)G_{T_2}^{-1}\big(x+1,y,s\big)\notag\\
&\cdot G_{T_1}\big(x+1,y,s\big)G_{T_2}(x,y,s)=\eta_{12}I_{m\times m},\\
&\notag G_{T_1}^{-1}(x,y,s)G_{P_x}(x,y,s)G_{T_1}^\ast(x,s-y,s)\\
&\cdot G_{P_x}^{-1}(x-1,y,s)=\eta_{x1}I_{m\times m},\\
&\notag G_{T_2}^{-1}(x,y,s)G_{P_x}(x,y,s)\big[G_{T_2}^{-1}(x,1+s-y,s)\big]^\ast\\
&\cdot G_{P_x}^{-1}(x,y-1,s)=\eta_{x2}I_{m\times m},\\
&\notag G_{T_1}^{-1}(x,y,s)G_{P_y}(x,y,s)\big[G_{T_1}^{-1}(2-s-x,y,s)\big]^\ast\\
&\cdot G_{P_y}^{-1}(x-1,y,s)=\eta_{y1}I_{m\times m},\\
&\notag G_{T_2}^{-1}(x,y,s)G_{P_y}(x,y,s)G_{T_2}^\ast(1-s-x,y,s)\\
&\cdot G_{P_y}^{-1}(x,y-1,s)=\eta_{y2}I_{m\times m},\\
&\notag G_{T_1}^{-1}(x,y,s)G_{\pxy}(x,y,s)G_{T_2}^\ast(y,x,1-s)\\
&\cdot G_{\pxy}^{-1}(x-1,y,s)=\eta_{xy1}I_{m\times m},\\
&\notag G_{T_2}^{-1}(x,y,s)G_{\pxy}(x,y,s)G_{T_1}^\ast(y,x,1-s)\\
&\cdot G_{\pxy}^{-1}(x,y-1,s)=\eta_{xy2}I_{m\times m},\\
&G_{P_x}(x,y,s)G_{P_x}^\ast(x,s-y,s)=\eta_x I_{m\times m},\\
&G_{P_y}(x,y,s)G_{P_y}^\ast(1-s-x,y,s)=\eta_x I_{m\times m},\\
&G_{\pxy}(x,y,s)G_{\pxy}^\ast(y,x,1-s)=\eta_{xy}I_{m\times m},\\
&\notag G_{P_x}(x,y,s)G^\ast_{P_y}(x,s-y,s)G_{P_x}(1-s-x,s-y,s)\\
&\cdot G_{P_y}^\ast(1-s-x,y,s)=\eta_{P_xP_y}I_{m\times m},\\
&\notag G_{\pxy}(x,y,s)G_{P_x}^\ast(y,x,1-s)G_\pxy(y,1-s-x,1-s)\\
&\cdot G^\ast_{P_y}(1-s-x,y,s)=\eta_{\pxy P_x}I_{m\times m}.
\end{eqnarray}
where again all $\eta$'s are $Z_m$ quantum numbers $\{e^{\imth 2\pi\frac km}|k=1,\cdots,m\}$. Note that $P_x,~P_y,~\pxy$ are all a anti-unitary symmetries and they are accompanied with complex conjugate $\mathcal{C}$. We can always choose a proper gauge so that $G_{T_2}(x,y,s)=G_{T_1}(x,0,s)=I_{m\times m}$. $\eta_{xy1}=1$ can also be fixed by certain gauge choice. After solving the above algebraic conditions one can see $\eta_{y1}=\eta_{x2}=\eta_{x}=\eta_y=\eta_{xy}=1$,~$\eta_{xy2}=\eta_{xy1}=1$,~$\eta_{y2}=\eta_{x1}^{-1}$. It turns out $\eta_{12},~\eta_{x1},~\eta_{P_xP_y},~\eta_{\pxy P_x}$ are the only four independent $Z_m$ quantum numbers. In the end we have
\begin{eqnarray}
&\label{psg:checkerboard}G_{T_2}(x,y,s)=I_{m\times m},~~~G_{T_1}(x,y,s)=\eta_{12}^yI_{m\times m},\\
&\notag G_{P_x}(x,y,s)=\eta_{12}^{sx}\eta_{x1}^xg_{x}(s),\\
&\notag G_{P_y}(x,y,s)=\eta_{x1}^{-y}g_{y}(s),~~~G_{\pxy}(x,y,s)=\eta_{12}^{xy}g_{xy}(s),\\
&\notag g_x(s)g_x^\ast(s)=g_y(s)g_y^\ast(s)=g_{xy}(s)g^\ast_{xy}(1-s)=1,\\
&\notag \big[g_x(s)g_y^\ast(s)\big]^2=\eta_{x1}^{-1}\eta_{P_xP_y},\\
&\notag g_{xy}(s)g_y^\ast(1-s)g_{xy}(1-s)g_x^\ast(s)=\eta_{P_{xy}P_x}.
\end{eqnarray}
where $g_x(s)$, $g_y(s)$ and $g_{xy}(s)$ are all $SU(m)$ matrices.

Different $Z_m$ mean-field states are given by gauge-inequivalent solutions of equations (\ref{psg:checkerboard}).

\subsection{Symmetry conditions on $Z_m$ mean-field states}

A representative of all mean-field amplitudes $M({\bf r}|{\bf r}^\prime)$ is given by ${\bf r}^\prime=(0,0,0)$ and we label independent mean-field amplitudes as
\begin{eqnarray}
[x,y,s]\equiv M(x,y,s|0,0,0)
\end{eqnarray}
All other symmetry-related mean-field amplitudes can be generated by $[x,y,s]$: \eg all 1st nearest neighbor (n.n.) mean-field amplitudes can be generated by $[0,0,1]$ through symmetry operations; all 2nd n.n. amplitudes can be generated by $[0,1,0]$ and $[1,0,0]$ \etc.

Using symmetry operations (\ref{symmetry:checkerboard}) it's straightforward to verify the following relations
\begin{eqnarray}
&\notag T_1^{x-1}T_2^yP_x P_y:~~[x,y,0]\rightarrow[
x,y,0]^\dagger\\
&T_2^{x}P_x:~~[0,x,0]\rightarrow[0,x,0]^\dagger\notag\\
&T_1^{-1}P_y:~~[0,x,0]\rightarrow[0,x,0]\notag\\
&T_1^{x-1}P_y:~~[x,0,0]\rightarrow[x,0,0]^\dagger\notag\\
&P_x:~~[x,0,0]\rightarrow[x,0,0]\notag
\end{eqnarray}
for mean-field amplitudes $[x,y,0]=M(x,y,0|0,0,0)$ between sites
within the same sublattice and
\begin{eqnarray}
&\notag T_1^xT_2^{-x}\pxy:~~[x,-x,1]\rightarrow[
x,-x,1]^\dagger\\
&T_1^{x}T_2^{x}P_x P_y\pxy:~~[x,x+1,1]\rightarrow[x,x+1,1]^\dagger\notag
\end{eqnarray}
for mean-field amplitudes $[x,y,1]=M(x,y,1|0,0,0)$ between sites
from different sublattices. The symmetry condition (\ref{mfH:symmetry}) now gives us constraints on the mean-field amplitudes $[x,y,s]$.

In the following we list the symmetry conditions on mean-field amplitudes up to 3rd nearest neighbor. For 1st n.n. $u_\alpha\equiv[0,0,1]$
\begin{eqnarray}
&\notag g_\pxy(0)u_\alpha^\ast g_\pxy^\dagger(1)=u_\alpha^\dagger.
\end{eqnarray}
for 2nd n.n. $u_{\beta x}\equiv[1,0,0]$
\begin{eqnarray}
\notag \eta_{x1}g_x(0)u_{\beta x}^\ast g_x^\dagger(0)=u_{\beta x},\\
\notag g_y(0)u_{\beta x}^\ast g_y^\dagger(0)=u^\dagger_{\beta x}.
\end{eqnarray}
for 2nd n.n. $u_{\beta y}\equiv[0,1,0]$
\begin{eqnarray}
\notag \eta_{x1}^{-1}\eta_{12}^{-1}g_y(0)u_{\beta y}^\ast g_y^\dagger(0)=u_{\beta y},\\
\notag g_x(0)u_{\beta y}^\ast g_x^\dagger(0)=u^\dagger_{\beta y}.
\end{eqnarray}
for 3rd n.n. $u_\gamma\equiv[1,1,0]$
\begin{eqnarray}
u_\gamma^\dagger=g_{y}(0)g^\ast_x(0)u_\gamma g^T_x(o)g^\dagger_y(0)
\end{eqnarray}

\section{Ground state degeneracy of $SU(m)_1$ Chern-Simons theory on a torus}\label{app:g.s.deg}

In this section we calculate the ground state degeneracy of a $SU(m)_1$ Chern-Simons theory
\begin{eqnarray}\label{action:su(m) cs}
&\notag\mathcal{L}_{SU(m)_1}=\frac1{4\pi}\epsilon_{\mu\nu\lambda}\text{Tr}\big(a_\mu\partial_\nu a_\lambda+\frac\imth3a_\mu a_\nu a_\lambda\big),~~a_\mu\in SU(m)\\
\end{eqnarray}
on a torus with $m$ being an odd integer. In \Ref{Wen1998} the ground state degeneracy of $SU(3)_q$ parton states on a torus is calculated. We follow their strategy\cite{Wen1999} and generalize it to the case of $SU(m)_1$.

After choosing a gauge $a_0\equiv0$, the classical configuration of gauge fields $a_\mu$ are constrained by the following condition $b=\epsilon_{ij}\partial_ia_j=0$. The gauge field configuration is fully determined by Wilson loop operators $U_{1,2}=\mathcal{P}\big[e^{\imth\oint_{c_{1,2}}a_\mu\text{d}x_\mu}\big]$ for the two non-contractible loops $c_{1,2}$ along $x_1$ and $x_2$ directions on the torus. These two loop operators commute with each other since $U_1U_2U_1^\dagger U_2^\dagger=1$ is the Wilson loop operator for a contractible loop. Therefore by a global $SU(m)$ transformation, we can see $U_{1,2}$ lie in the maximal Abelian subgroup of $SU(m)$, which is generated by the Cartan subalgebra of Lie algebra $su(m)$. We choose the generator of its Cartan subalgebra to be $m\times m$ matrices
\begin{eqnarray}
&\label{generator: Cartan subalgebra of su(m)} g^1=\text{Diag}(1,-1,0,\cdots,0),\\
&\notag g^2=\text{Diag}(1,1,-2,0,\cdots,0),\\
&\notag g^3=\text{Diag}(1,1,1,-3,0,\cdots,0),\\
&\notag \cdots\cdots\cdots\\
&\notag g^{m-1}=\text{Diag}(1,\cdots,1,-m-1).
\end{eqnarray}
Therefore the gauge field configuration is given by
\begin{eqnarray}
\notag a_0=0,~~~a_j=\sum_{I=1}^{m-1}a^I_jg^I.
\end{eqnarray}
Plugging this into (\ref{action:su(m) cs}) we have
\begin{eqnarray}\label{action:su(m) cs1}
\notag\mathcal{L}_{SU(m)_1}=\frac1{4\pi}\epsilon_{\mu\nu\lambda}\Big(\sum_{I=1}^{m-1}I(I+1)a^I_\mu\partial_\nu a^I_\lambda\Big)
\end{eqnarray}
In this way the original Chern-Simons theory of non-Abelian $SU(m)$ gauge fields is reduced to $(m-1)$ different Chern-Simons theory of $U(1)$ gauge fields $a_\mu^I,~I=1,\cdots,m-1$. In addition to these Abelian $U(1)$ gauge structures, there are discrete $SU(m)$ gauge transformations generated by $W_i\in SU(m)$ which leaves this maximial Abelian subgroup invariant. The low-energy degrees of freedom are described\cite{Elitzur1989,Polychronakos1990} by vectors $\vec u(t)$ and $\vec v(t)$ with
\begin{eqnarray}
a_1^I(x_1,x_2,t)=\frac{2\pi}{L_1}u_I(t),~~~a_2^I(x_1,x_2,t)=\frac{2\pi}{L_2}v_I(t).
\end{eqnarray}
on a $L_1\times L_2$ torus. The effective action becomes
\begin{eqnarray}
\mathcal{L}=2\pi\sum_{I=1}^{m-1}I(I+1)v_I\dot{u}_I
\end{eqnarray}
This immediately leads to the following canonical commutation relations
\begin{eqnarray}
[u_I,v_J]=\frac{\imth}{2\pi}\cdot\frac1{I(I+1)}\delta_{I,J}
\end{eqnarray}
In other words, the conjugate momenta of coordinate $u_I$ is $2\pi I(I+1)v_I$. Due to uncertainty principle, they cannot be fixed simultaneous in a quantum state.

There are large gauge transformations $\{U^I_i=\exp\big(\imth2\pi x_ig^I/L_i\big),~i=1,2\}$ acting on the partons, which leaves both the electron operators and the physical electron states invariant. The conjugate variables transform in the following way under the large gauge transformations
\begin{eqnarray}
u_I\sim u_I+1,~~~v_I\sim v_I+1.\notag
\end{eqnarray}
therefore they live on a torus of size $1\times1$. As a result the wavefunctions can be written as
\begin{eqnarray}
\psi(\vec u)=\sum_{\vec n}c_{\vec n}e^{2\pi\imth\vec n\cdot\vec u}
\end{eqnarray}
where $\vec n$ is a $(m-1)$-dimensional vector of integers. Since the conjugate momentum of $u_I$ is $2\pi I(I+1)v_I$ we have
\begin{eqnarray}
\psi(\vec u)\sim\sum_{\vec n}c_{\vec n}\prod_{I=1}^{m-1}\delta_{I(I+1)v_I,n_I}
\end{eqnarray}
The large gauge transformation $v_I\sim v_I+1$ enforces the periodic condition $c_{\vec n}=c_{\vec n+\delta\vec n}$ where $\delta\vec n_I=I(I+1)Z_I,~Z_I\in\mathbb{Z}$.

Aside from the large gauge transformations $U^I_i$, there are other large gauge transformations under which
\begin{eqnarray}
&\vec{u}\equiv\begin{pmatrix}u_1\\u_2\\ \cdots\\u_{m-1}\end{pmatrix}\sim\label{cs_other_large_g.t.}\\
&\vec u+\begin{bmatrix}{-1}&1&1&\cdots&1\\0&-2&1&\cdots&1\\0&0&-3&\cdots&1\\
\cdots&\cdots&\cdots&\cdots&\cdots\\0&0&0&\cdots&-m-1\end{bmatrix}^{-1}\begin{pmatrix}d_2\\d_3\\ \cdots\\d_{m}\end{pmatrix}\notag
\end{eqnarray}
and similar relations hold for $\vec v$. Here $\{d_I,~I=2,\cdots,m\}$ are arbitrary integers and $d_1=-\sum_{I=2}^md_I$. The partons transform as $f_\alpha(x_1,x_2)\rightarrow\exp(\imth2\pi x_1d_\alpha/L_1)f_\alpha(x_1,x_2)$. As mentioned earlier, in addition to the large gauge transformations, there are discrete $SU(m)$ gauge transformations essentially generated by permutation between the $m$ species of partons. For example, the permutation $P_{12}$ transforms the partons as $f_1\leftrightarrow f_2$ and thus we have
\begin{eqnarray}
P_{12}:~~~&\vec u=(u_1,u_2,\cdots)^T\sim(-u_1,u_2,\cdots)^T\notag\\
&\vec v=(v_1,v_2,\cdots)^T\sim(-v_1,v_2,\cdots)^T\notag
\end{eqnarray}
There are in total $m(m-1)/2$ permutations $\{P_{\alpha\beta}|\alpha<\beta\}$ that leaves the maximal Abelian subgroup of $SU(m)$ invariant. All these gauge transformations give us further equivalence conditions on $\vec u,\vec v$, and thus on coefficients $c_{\vec n}$. In the end, the number of independent coefficients $c_{\vec n}$ equals the ground state degeneracy of $SU(m)_1$ Chern-Simons theory on a torus. With the help of computer programming it's easy to check the ground state degeneracy of $SU(m)_1$ Chern-Simons theory on a torus is $m$, and they correspond to the following (gauge-inequivalent) gauge field configurations
\begin{eqnarray}
&\notag v_I=0,~(I=1,2,\cdots,m-2);\\
&v_{m-1}=\frac km,~~~k=1,2,\cdots,m.
\end{eqnarray}
Physically they correspond to inserting $2\pi k/m$ flux through the hole along $x_1$ direction.
An explicit example of $SU(3)_1$ Chern-Simons theory is given in \Ref{Wen1998}.

\section{Symmetry analysis of time-reversal-invariant FTI states}\label{app:psg:fti:checkerboard}

In this section we use projected symmetry group (PSG) to analyze symmetric mean-field ansatz realized in a time-reversal invariant lattice model in $SU(m)$ parton construction. We require the corresponding electron state to preserve both time reversal symmetry $T$ and all lattice symmetries.

In the following we use the time-reversal-invariant checkerboard lattice model\cite{Sun2011,Neupert2011a} as an example. The discussion is completely general and can be applied to any lattice model. The symmetry group is generated by lattice symmetry (\ref{symmetry:checkerboard}) together with time reversal symmetry:
\begin{eqnarray}\label{sym:T}
\bst:~~~(x,y,s,\sigma)\rightarrow(x,y,s,1-\sigma)
\end{eqnarray}
where $\sigma=0/1$ represent the electron spin $\uparrow/\downarrow$. Note that here $P_x, P_y$ and $\pxy$ are the combination of spatial $\pi$ rotations (along $\hat x$, $\hat y$ and $\hat x+\hat y$-axis) and spin rotations. So all these three generators flip spin: $\sigma\rightarrow1-\sigma$. They are not accompanied with time-reversal operation anymore and they are now unitary symmetries. Time reversal itself is still an anti-unitary symmetry realized by complex conjugation operation $\mathcal{C}$. The algebra structure of this symmetry group is determined by (\ref{algebra:checkerboard_fqh}) together with
\begin{eqnarray}
&\bst^2=\bse;\\
&\notag g^{-1}\bst^{-1}g\bst=\bse,\\
&\notag g=T_{1,2},~P_x,~P_y,~\pxy.
\end{eqnarray}
Consider a $IGG=Z_m^\uparrow\times Z_m^\downarrow$ mean-field ansatz and we have the following algebraic conditions for the gauge transformations $G_U(x,y,s,\sigma)$ associated with $U=\bst,T_{1,2}$
\begin{eqnarray}
&\notag G_{T_1}^{-1}(x,y,s,\sigma)G_\bst(x,y,s,\sigma)G^\ast_{T_1}(x,y,s,1-\sigma)\\
&G_\bst^{-1}(x-1,y,s,\sigma)=\eta_{\bst1},\label{psg:tfi:T1T}\\
&G_{T_2}^{-1}(x,y,s,\sigma)G_\bst(x,y,s,\sigma)G^\ast_{T_2}(x,y,s,1-\sigma)\notag\\
&G_\bst^{-1}(x,y-1,s,\sigma)=\eta_{\bst2}.\label{psg:tfi:T2T}
\end{eqnarray}
since $T_{1,2}^{-1}\bst T_{1,2}\bst^{-1}=\bse$. Here $\eta_{\bst1},~\eta_{\bst2}$ are all $Z_m$ quantum numbers $\{e^{\imth\frac{2\pi a}m}|a=1,\cdots,m\}$. As shown in Appendix \ref{app:psg:checkerboard}, one can always choose a certain gauge so that
\begin{eqnarray}
G_{T_2}(x,y,s,\sigma)=I_{m\times m},~~G_{T_1}(x,y,s,\sigma)=\eta_{12}(\sigma)I_{m\times m}.\notag
\end{eqnarray}
where $\eta_{12}(\sigma),~\sigma=0/1$ are also $Z_3$ quantum numbers. From (\ref{psg:tfi:T1T}) and (\ref{psg:tfi:T2T}) one can solve out
\begin{eqnarray}
&\notag G_\bst(x,y,s,\sigma)=\eta_{\bst1}^x\eta_{\bst2}^yg_\bst(s,\sigma),\\
&\eta_{12}(\uparrow)\eta_{12}(\downarrow)=1.\label{G_T}
\end{eqnarray}
Therefore when $IGG=Z_m^\uparrow\times Z_m^\downarrow$ for the parton mean-field state, we can simply choose $\eta_{12}(\downarrow)=\eta_{12}^{-1}(\uparrow)$ to preserve the time-reversal symmetry. However, when there are spin mixing terms in the mean-field parton Hamiltonian, the gauge symmetry is broken further down to $IGG=Z_m$ and we must have $\eta_{12}(\uparrow)=\eta_{12}(\downarrow)$. From (\ref{G_T}) we know in order to preserve time-reversal symmetry, we must have $\eta_{12}(\uparrow)=\eta_{12}(\downarrow)=1$ in the case $IGG=Z_m$. As a result there is no way to insert $2\pi/m$ flux into each unit cell in the parton mean-field ansatz if we require time-reversal symmetry and lattice translation symmetry. Therefore the technique of enlarging unit cell by $m$ times in order to construct a gapped state with filling fraction $\nu=1/m$ failed. This suggests that a gapped FTI state with gauge structure $IGG=Z_m$ at filling fraction $1/m$ would break either time reversal or lattice symmetries.

\section{The honeycomb lattice model: four different examples of spin-polarized FCI states at $\nu=1/3$}\label{app:example:honeycomb}

Haldane's model on honeycomb lattice\cite{Haldane1988} has been shown to support nearly flat bands with non-Zero Chern numbers\cite{Neupert2011}. Its symmetry group is shown in Appendix \ref{app:sym:honeycomb} and FIG. \ref{fig:lattice}. We label a lattice site by coordinate $(x,y,s)$ as explained in Appendix \ref{app:sym:honeycomb}. By $SU(3)$ parton construction, we use PSG to classify different spin-polarized $Z_3$ mean-field ansatz as shown in Appendix \ref{app:psg:honeycomb}. In the following we show two mean-field states belonging to different universality classes. They correspond to two gauge-inequivalent solutions to (\ref{psg:honeycomb}). Their two $SU(3)$ parent states also have different PSGs and thus distinct. We show mean-field amplitudes up to NNN in FIG. \ref{fig:hc123}.

\subsection{FCI state HC1 with $\sigma_{xy}=1/3$:~the $Z_3$ state and its parent $SU(3)$ state}

\begin{figure}
 \includegraphics[width=0.4\textwidth]{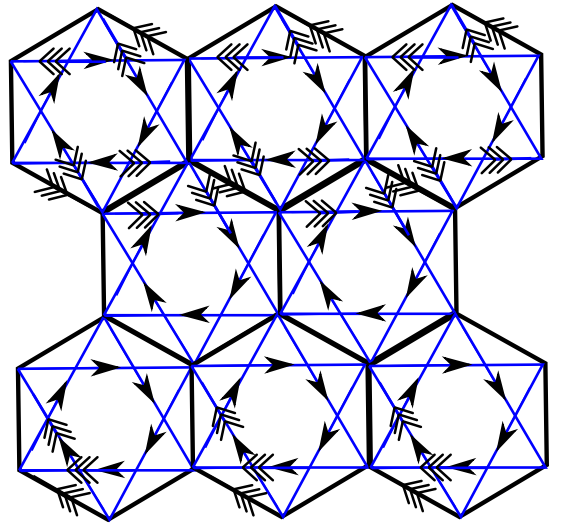}
\caption{(color online) Mean-field ansatz of the parent $SU(3)$ parton states associated with FCI states HC1 and HC3 on honeycomb lattice. The solid line for NN bonds represents real hopping amplitude $\alpha$. NNN bonds represents complex hopping amplitude $\beta$ along direction of the arrow. The triple arrow means the original hopping amplitude along its direction should be multiplied by a phase factor $\eta_{12}$. Note in the mean-field ansatz the lattice translation along $\vec a_2$ direction is explicitly broken by flux insertion and the unit cell is tripled. However the lattice translation symmetry is preserved in the corresponding electron states after projection (\ref{projection:1/m}).}\label{fig:hc123}
\end{figure}

In spin-polarized $Z_3$ FCI state HC1 the gauge transformations $G_U(x,y,s)$ associated with lattice symmetry $U$ are listed below:
\begin{eqnarray}
&\notag G_{T_1}(x,y,s)=I_{3\times3},~~~G_{T_2}(x,y,s)=\eta_{12}^xI_{3\times3},\\
&\notag G_\bss(x,y,s)=\eta_{12}^{-x-y(y+1)/2}I_{3\times3},\\
&\notag G_\cs(x,y,s)=\eta_{12}^{xy+x(x-1)/2}I_{3\times3}
\end{eqnarray}
where $\eta_{12}=\exp(\imth2\pi/3)$. As shown in Appendix \ref{app:psg:honeycomb} the symmetry allowed mean-field amplitudes are:

(\Rmnum{1}) For nearest neighbor (NN) amplitude $u_\alpha\equiv M(0,0,1|0,0,0)$
\begin{eqnarray}
u_\alpha=u_\alpha^T=u_\alpha^\ast
\end{eqnarray}
in other words $u_\alpha$ can be any real symmetric $3\times3$ matrix. All other NN amplitudes can be generated from $u_\alpha$ by symmetry operations through (\ref{mfH:symmetry}).

(\Rmnum{2}) For next nearest neighbor (NNN) amplitude $u_\beta\equiv M(0,1,0|0,0,0)$
\begin{eqnarray}
u_\beta=u_\beta^T
\end{eqnarray}
\ie $u_\beta$ can be any symmetric $3\times3$ matrix. All other NNN mean-field amplitudes can be generated from $u_\alpha$ by symmetry operations through (\ref{mfH:symmetry}).

\begin{figure}
 \includegraphics[width=0.45\textwidth]{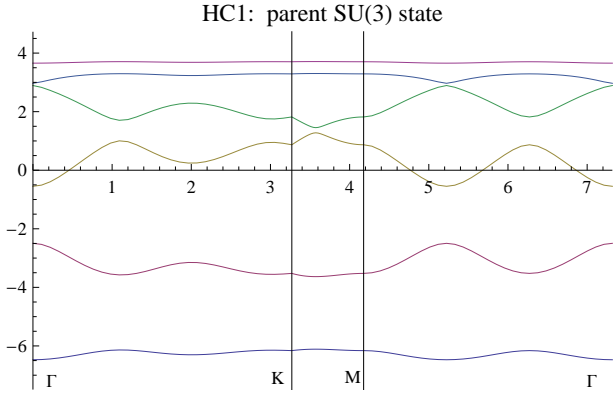}
\caption{(color online) The parton band structure of the parent $SU(3)$ ansatz (\ref{mfH:su(m)}) of HC1 state. Each band is 3-fold degenerate, corresponding to 3 parton flavors $f_{1,2,3}$. We plot the dispersion along $\Gamma\rightarrow K\rightarrow M\rightarrow\Gamma$ as shown in FIG. \ref{fig:bz}(a). Hopping parameters are chosen as $\alpha=-1$ for NN,~$\beta=-0.22\exp[0.2\imth]$ for NNN. The Chern numbers of the 6 bands are $\{1,-2,-2,1,1,1\}$ in a bottom-up order.}\label{fig:hc1_band}
\end{figure}

Its parent $SU(3)$ mean-field state includes both NN and NNN hopping terms $u_\alpha=\alpha I_{3\times3},~u_\beta=\beta I_{3\times3}$, as shown in FIG. \ref{fig:hc123}. With hopping parameters chosen as $\alpha=-1$, $\beta=-0.22\exp[0.2\imth]$, the parton band structure is shown in FIG. \ref{fig:hc1_band}. In the bottom-up order the Chern numbers are $\{1,-2,-2,1,1,1\}$ for the six bands (each is 3-fold degenerate) and these band structures persist in a large parameter range. As we add small terms mixing different species of partons to break the gauge symmetry down to $Z_3$, we have 18 non-degenerate parton bands and the Chern number for the lowest three bands are all $+1$. And the lowest three parton bands are well separated from the other 15 bands by a large gap. As a result the $Z_3$ FCI state we obtained by filling the lowest three bands has a Hall conductance $3\cdot(1/3)^2=1/3$ in the unit of $e^2/h$ since each parton carries electric charge $e/3$.

\subsection{FCI state HC2 with $\sigma_{xy}=1/3$:~the $Z_3$ state and its parent $SU(3)$ state}

In $Z_3$ FCI state HC2 the gauge transformations $G_U(x,y,s)$ associated with lattice symmetry $U$ are listed below:
\begin{eqnarray}
&\notag G_{T_1}(x,y,s)=I_{3\times3},~~~G_{T_2}(x,y,s)=\eta_{12}^xI_{3\times3},\\
&\notag G_\bss(x,y,s)=\eta_{12}^{-x-y(y+1)/2}I_{3\times3},\\
&\notag G_\cs(x,y,0)=\eta_{12}^{xy+x(x-1)/2}I_{3\times3},\\
&\notag G_\cs(x,y,1)=\eta_{12}^{xy+x(x-1)/2}\text{Diag}\big\{e^{\imth2\pi/9},e^{\imth2\pi/9},e^{-\imth4\pi/9}\big\}.
\end{eqnarray}
where $\eta_{12}=\exp(\imth2\pi/3)$.

As shown in Appendix \ref{app:psg:honeycomb} the symmetry allowed mean-field amplitudes are:

(\Rmnum{1}) For nearest neighbor (NN) amplitude $u_\alpha\equiv M(0,0,1|0,0,0)$
\begin{eqnarray}
u_\alpha=\begin{bmatrix}e^{\imth\pi/9}\alpha_{11}&e^{\imth\pi/9}\alpha_{12}&e^{-\imth\pi/18}\alpha_{13}\\
e^{\imth\pi/9}\alpha_{12}&e^{\imth\pi/9}\alpha_{22}&e^{-\imth\pi/18}\alpha_{23}\\
e^{-\imth\pi/18}\alpha_{13}&e^{-\imth\pi/18}\alpha_{23}&e^{-\imth2\pi/9}\alpha_{33}\end{bmatrix}
\end{eqnarray}
where all $\alpha_{ij}$ are real parameters. All other NN amplitudes can be generated from $u_\beta$ by symmetry operations through (\ref{mfH:symmetry}).

(\Rmnum{2}) For next nearest neighbor (NNN) amplitude $u_\beta\equiv M(0,1,0|0,0,0)$
\begin{eqnarray}
u_\beta=\begin{bmatrix}\beta_{11}&\beta_{12}&\beta_{13}e^{\imth\pi/3}\\
\beta_{12}&\beta_{22}&\beta_{23}e^{\imth\pi/3}\\
\beta_{13}e^{-\imth\pi/3}&\beta_{23}e^{-\imth\pi/3}&\beta_{33}
\end{bmatrix}
\end{eqnarray}
where all $\beta_{ij}$ are complex parameters. All other NNN mean-field amplitudes can be generated from $u_\beta$ by symmetry operations through (\ref{mfH:symmetry}).

\begin{figure}
 \includegraphics[width=0.45\textwidth]{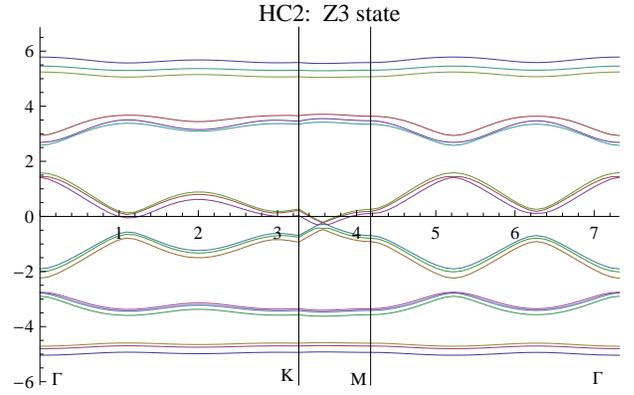}
\caption{(color online) The parton band structure of a $Z_3$ mean-field ansatz: HC2 state. There are 18 non-degenerate parton bands. We plot the dispersion along $\Gamma\rightarrow K\rightarrow M\rightarrow\Gamma$ as shown in FIG. \ref{fig:bz}(a). The Chern numbers of the lowest 6 bands are $\{1,1,1,-2,-2,-2\}$ in a bottom-up order.}\label{fig:hc2_z3}
\end{figure}

Its parent $SU(3)$ mean-field state include only NNN hopping terms, \ie $u_\alpha=0$ and $u_\beta=\beta I_{3\times3}$. It is demonstrated by the NNN bonds shown in FIG. \ref{fig:hc123}. Apparently this mean-field ansatz is very different from HC1 state since its parent $SU(3)$ state doesn't allow NN hopping terms. We choose the following hopping parameters: $\alpha_{11}=\alpha_{22}=\alpha_{33}=2$, $\alpha_{12}=0.07,~\alpha_{13}=0.04,~\alpha_{23}=0.05$ and $\beta_{11}=\beta_{22}=\beta_{33}=0.1\exp[0.2\imth]$,~$\beta_{12}=0.03\exp[\imth0.7],~\beta_{13}=0.07\exp[\imth0.2],~\beta_{23}=0.06\exp[\imth1.9]$. We find that in a very large parameter range the Chern number of the lowest three parton bands remains to be $+1$ and they are separated from the other bands by a large gap, as shown in FIG. \ref{fig:hc2_z3}. By filling these 3 lowest bands we obtain the $Z_3$ FCI state HC2 which also has Hall conductivity $1/3$ in the unit of $e^2/h$.

\section{Two examples of non-Abelian spin-polarized $SU(3)$ FCI states}\label{app:example}

As discussed in section \ref{EFT}, by partially filling a topological flat band with Chern number $C>1$, it might be possible to realize non-Abelian FCI states. Here we show two examples of spin-polarized $SU(3)$ FCI states with $\nu=1/3$, one (labeled as HC3) on honeycomb lattice and the other (labeled as CB3) on checkerboard lattice. In particular these states are likely to be non-Abelian FCI states associated with $SU(3)_2$ Chern-Simons theory\cite{Wen1998} since the each parton species fills a lowest band with Chern number $+2$. Their Hall conductance is $\sigma_{xy}=\frac23$ in the unit of $e^2/h$. By braiding non-Abelian quasiparticles in these FCIs one can carry out universal quantum computations\cite{Freedman2002a,Nayak2008}.

\subsection{Honeycomb lattice: $SU(3)$ FCI state HC3 with $\sigma_{xy}=2/3$ and $\nu=1/3$}

\begin{figure}
 \includegraphics[width=0.45\textwidth]{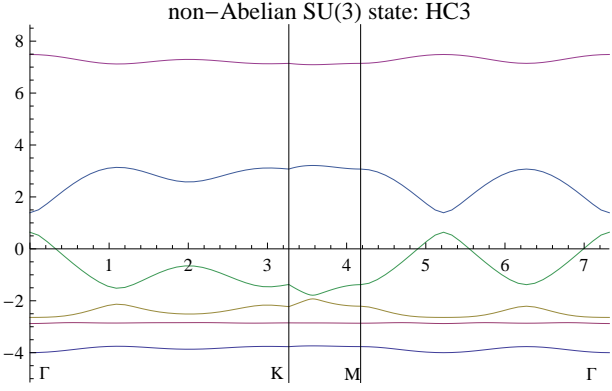}
\caption{(color online) The parton band structure of HC3 state, a non-Abelian $SU(3)$ FCI state on honeycomb lattice. Each band is 3-fold degenerate, corresponding to 3 parton flavors $f_{1,2,3}$. We plot the dispersion along $\Gamma\rightarrow K\rightarrow M\rightarrow\Gamma$ as shown in FIG. \ref{fig:bz}(a). Its mean-field ansatz are shown in FIG. \ref{fig:hc123} with $\eta_{12}=\exp(-\imth2\pi/3)$. Hopping parameters are chosen as $\alpha=-1$ for NN and $\beta=-0.22\exp[0.2\imth]$ for NNN. The Chern numbers of the 6 bands are $\{2,2,-1,-4,2,-1\}$ in a bottom-up order.}\label{fig:hc3_band}
\end{figure}

The $SU(3)$ FCI state HC3 is an example of non-Abelian states on the honeycomb lattice. It includes both NN and NNN hopping terms $u_\alpha=\alpha I_{3\times3},~u_\beta=\beta I_{3\times3}$ in its mean-field ansatz, as shown in FIG. \ref{fig:hc123} with $\eta_{12}=\exp(-\imth2\pi/3)$.  It is different from the parent $SU(3)$ state of HC1, since in HC1 we insert $-2\pi/3$ flux per unit cell and here we insert $2\pi/3$ flux in HC3 state to enlarge the unit cell. With hopping parameters chosen as $\alpha=1$, $\beta=0.35\exp[0.2\imth]$, its parton band structure is shown in FIG. \ref{fig:hc3_band}. In the bottom-up order the Chern numbers are $\{2,2,-1,-4,2,-1\}$ for the six bands (each is 3-fold degenerate) and these band structures persist in a large parameter range. And the lowest parton band is separated from the other 5 bands by a sizable gap. As a result the $SU(3)$ FCI state HC3 we obtained by filling the lowest 3-fold-degenerate band has a Hall conductance $\sigma_{xy}=3\cdot2\cdot(1/3)^2=2/3$ in the unit of $e^2/h$, since each parton carries electric charge $e/3$.

\subsection{Checkerboard lattice: $SU(3)$ FCI state CB3 with $\sigma_{xy}=2/3$ and $\nu=1/3$}

\begin{figure}
 \includegraphics[width=0.45\textwidth]{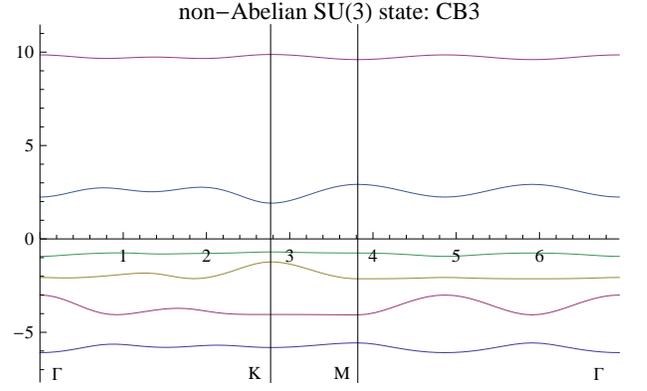}
\caption{(color online) The parton band structure of CB3 state, a non-Abelian $SU(3)$ FCI state on checkerboard lattice. Each band is 3-fold degenerate, corresponding to 3 parton flavors $f_{1,2,3}$. We plot the dispersion along $\Gamma\rightarrow K\rightarrow M\rightarrow\Gamma$ as shown in FIG. \ref{fig:bz}(b). Its mean-field ansatz are shown in FIG. \ref{fig:cb23} with $\eta_{12}=\exp(\imth2\pi/3)$. Hopping parameters are chosen as $\alpha=e^{\imth\pi/12}$ for NN,~$\beta_x=0.6=\beta_y$ for NNN and $\gamma=0.5$ for NNNN. The Chern numbers of the 6 bands are $\{2,2,-1,-4,2,-1\}$ in a bottom-up order.}\label{fig:cb3_band}
\end{figure}

The $SU(3)$ FCI state CB3 is an example of non-Abelian FCI states on checkerboard lattice. It includes NN hopping $u_\alpha=\alpha I_{3\times3}$, NNN hopping $u_{\alpha x}=e^{-\imth\pi/3}\beta_x I_{3\times3},~u_{\alpha y}=\beta_y I_{3\times3}$ and NNNN hopping $u_{\gamma}=\gamma I_{3\times3}$ in its mean-field ansatz, as shown in FIG. \ref{fig:cb23} with $\eta_{12}=\exp(\imth2\pi/3)$. It is different from the parent $SU(3)$ state of CB2, since in CB2 $2\pi/3$ flux are inserted into each unit cell while here we insert $-2\pi/3$ flux for CB3. Choosing parameters $\alpha=e^{\imth\pi/12}$,~$\beta_x=0.6=\beta_y$ and $\gamma=0.5$ we have the Chern numbers of the 6 bands for each parton species as $\{2,2,-1,-4,2,-1\}$ and the lowest band is well separated from other bands, as shown in FIG. \ref{fig:cb3_band}. This qualitative band structure persists for a large parameter range. Each band of a $SU(3)$ parton ansatz is 3-fold degenerate, corresponding to the 3 parton flavors $f_{1,2,3}$. By filling the 3-fold-degenerate lowest band (with Chern number $+2$) we obtain a $SU(3)$ FCI state CB3 whose Hall conductivity is $\sigma_{xy}=2/3$ in the unit of $e^2/h$.


\end{document}